\documentclass[conference,letterpaper]{IEEEtran}

\ifCLASSINFOpdf
  \usepackage{xcolor}
  \usepackage[pdftex]{graphicx}
\else
\fi

\usepackage{amsmath}
\usepackage{bbold} %

\usepackage[allowmove]{url}

\usepackage[inline]{enumitem}
\usepackage[hidelinks,%
draft=false,bookmarks=true   %
]{hyperref} 
\usepackage[noabbrev,capitalise]{cleveref}
\usepackage{booktabs}

\usepackage{multirow,makecell}
\usepackage{pgfplotstable}
\pgfplotsset{compat=1.18} 

\usepackage{tabularx}   %
\usepackage{etoolbox} %
\usepackage{calc}
\usepackage{adjustbox}
\usepackage{bbm}

\usepackage{siunitx}
\usepackage{listings} %

\usepackage[backend=biber, abbreviate=true, dateabbrev=true, alldates=year, isbn=false, doi=true, urldate=comp, url=false, maxbibnames=9, maxcitenames=2, backref=false, style=numeric-comp, language=american, giveninits=true, sortcites=true, sorting=none]{biblatex}

\usepackage{citation-cleaner}

\lstdefinestyle{prompt}{
    aboveskip=3pt,
    abovecaptionskip=3pt,
    xleftmargin=2pt, 
    xrightmargin=0pt,
    framexleftmargin=2pt,
    language={}, 
    morestring=[s]{<}{>},%
    stringstyle=\color{blue},
    basicstyle=\scriptsize\ttfamily,
    showstringspaces=false,
    breaklines=true,
    frame=lines,
    rulecolor=\color{black!20},
    backgroundcolor=\color{black!3},
}

\newcommand{\head}[1]{\par\noindent\textbf{#1:}\space}

\newenvironment{cveanalysis}[5]{%
    \begin{table}\centering
        \caption{Analysis of results for: \href{https://nvd.nist.gov/vuln/detail/#1}{#1}}\label{cve:#1}
        \begin{adjustbox}{width=.99\columnwidth}
        \begin{tabular}{p{.99\columnwidth}}
        \toprule
        \textbf{Description:} {#2} \\
        \ifstrempty{#3}{}{%
            \textbf{Exploitation Technique:} {#3}\\
        }%
        \ifstrempty{#4}{}{%
            \textbf{Primary Impact:} {#4}\\
        }%
        \ifstrempty{#5}{}{%
            \textbf{Secondary Impact:} {#5}\\
        }%
        \midrule
        \textbf{Analysis:}%
}{%
        \\\bottomrule
        \end{tabular}
        \end{adjustbox}
    \end{table}
}

\pgfplotstableset{
    string replace*={foobar}{quix}
}

\newcommand{\previousvalue}{}

\pgfplotstableset{
    col sep=comma,
    string type,
    begin table=\begin{tabular},
    end table=\end{tabular},
    column type=l,
    font=\footnotesize,
    every last row/.style={after row=\bottomrule},
    replace mapping types/.style={%
        columns/#1/.append style={ 
            string replace*={exploitation-technique}{exploitation},
            string replace*={primary-impact}{primary},
            string replace*={secondary-impact}{secondary},
        },
    },
    replace models/.style={%
        columns/#1/.append style={ 
            string replace*={llama3.3-70b-basic}{\makebox[\widthof{\approach{} (Llama3.3-70B)}][l]{ICL (Llama3.3-70B)}},
            string replace*={llama3.3-70b-combined}{\approach{} (Llama3.3-70B)},
            string replace*={gpt-4o-mini-basic}{ICL (GPT-4o-mini)},
            string replace*={gpt-4o-mini-combined}{\approach{} (GPT-4o-mini)}
        },
    },
    replace features/.style={%
        columns/#1/.append style={ 
            string replace*={ACVSSCWE}{\textsc{a}+\textsc{cvss}+\textsc{cwe}},
            string replace*={ACVSS}{\textsc{a}+\textsc{cvss}},
            string replace*={A}{\textsc{a}},
        },
    },
    replace mapping methods/.style={%
        columns/#1/.append style={ 
            string replace*={two-step-approach}{\makebox[\widthof{exploitation technique}][l]{In-Context Learner}}, 
            string replace*={methodology-combine}{\makebox[\widthof{exploitation technique}][l]{MM combined}}, 
            string replace*={vulnerability-type}{vulnerability type},
            string replace*={affected-object}{affected object types},
            string replace*={tactic-technique}{tactic},
            string replace*={exploitation-technique}{exploitation technique},
            string replace*={vulnerability-type}{vulnerability type},
        },
    },
    type siunitx/.style={%
        columns/#1/.append style={ string type, column type=S }
    },
    type right aligned/.style={%
        columns/#1/.append style={ string type, column type=r }
    },
    suppress duplicates/.style={
        columns/#1/.append style={
            string type,
            postproc cell content/.append code={%
                \pgfkeysgetvalue{/pgfplots/table/@cell content}\value
                \ifx\previousvalue\value
                    \pgfkeysalso{@cell content={}}%
                \else
                    \xdef\previousvalue{\value}%
                \fi
            }
        }
    }
}
\newcommand{\changetabcolsep}[1]{%
    \setlength{\tabcolsep}{#1}
    \setlength{\cmidrulekern}{\tabcolsep} %
}
\changetabcolsep{3pt}

\newcommand{\headingcell}[1]{\multirow{2}{*}[-2pt]{\makecell[l]{#1}}}

\newcommand{\citeboth}[1]{\citeauthor{#1}~\cite{#1}}

\newcommand{\attack}{\textsc{Att\&ck}}
\newcommand{\itattack}{A{\footnotesize TT}\&{\footnotesize CK}}  %
\newcommand{\approach}{\textsc{Triage}} 
\newcommand{\itapproach}{T{\footnotesize RIAGE}} %
\newcommand{\methodology}{CMM} %

\Crefname{appendix}{Appendix}{Appendices}

\setlist[enumerate,1]{%
    ref=\thesubsection-\arabic*%
}
\crefname{enumi}{}{}
\Crefname{enumi}{}{}

\makeatletter%
\begin{document}
\thispagestyle{plain}
\pagestyle{plain}
\makeatletter
\def\ps@IEEEtitlepagestyle{%
  \def\@oddfoot{\mycopyrightnotice}%
  \def\@evenfoot{}%
}
\def\mycopyrightnotice{%
  \hspace*{3mm}\includegraphics[width=2cm]{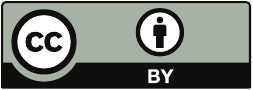}%
  \hspace*{2mm}\raisebox{2.5mm}{%
          \parbox{\columnwidth}{\footnotesize This work is licensed under a Creative Commons \\ Attribution 4.0 International (CC BY 4.0) license.}%
          \hspace*{-69pt}\mbox{\thepage}%
  }%
  \gdef\mycopyrightnotice{}%
}
\makeatother

\title{A Systematic Approach to Predict the Impact of Cybersecurity Vulnerabilities Using LLMs}

\author{%
  \IEEEauthorblockN{Anders M{\o}lmen H{\o}st}
  \IEEEauthorblockA{Simula \& University of Oslo\\
  Oslo, Norway \\
  andersmh@simula.no}
\and
  \IEEEauthorblockN{Pierre Lison}
  \IEEEauthorblockA{Norwegian Computing Center\\
  Oslo, Norway \\
  plison@nr.no}
\and
  \IEEEauthorblockN{Leon Moonen}
  \IEEEauthorblockA{Simula Research Laboratory\\
  Oslo, Norway \\
  leon.moonen@computer.org}
}

\maketitle

\noindent\begin{abstract}
Vulnerability databases, such as the National Vulnerability Database (NVD), offer detailed descriptions of Common Vulnerabilities and Exposures (CVEs), 
but often lack information on their real-world impact, such as the tactics, techniques, and procedures (TTPs) that adversaries may use to exploit the vulnerability.
However, manually linking CVEs to their corresponding TTPs is a challenging and time-consuming task, and the high volume of new vulnerabilities published annually makes automated support desirable.

This paper introduces \approach{}, a two-pronged automated approach that uses Large Language Models (LLMs) to map CVEs to relevant techniques from the \attack{} knowledge base.
We first prompt an LLM with instructions based on MITRE's CVE Mapping Methodology to predict an initial list of techniques. 
This list is then combined with the results from a second LLM-based module that uses in-context learning to map a CVE to relevant techniques. 
This hybrid approach strategically combines rule-based reasoning with data-driven inference. 
Our evaluation reveals that in-context learning outperforms the individual mapping methods, 
and the hybrid approach improves recall of exploitation techniques. 
We also find that GPT-4o-mini performs better than Llama3.3-70B on this task. 
Overall, our results show that LLMs can be used to automatically predict the impact of cybersecurity vulnerabilities and \approach{} makes the process of mapping CVEs to \attack{} more efficient.
\end{abstract}

\begin{IEEEkeywords}
vulnerability impact,
CVE, %
\attack{} techniques,
large language models,
automated mapping.
\end{IEEEkeywords}

\section{Introduction}\label{sec:introduction}

\noindent
Vulnerability databases, such as the National Vulnerability Database (NVD),\footnote{~\url{https://nvd.nist.gov}} help cybersecurity professionals keep track of publicly disclosed vulnerabilities (CVEs),\footnote{~\url{https://cve.org}} along with their corresponding labels. 
These labels include vulnerability descriptions, severity scores (CVSS), weakness classifications (CWEs),\footnote{~\url{https://cwe.mitre.org}} and affected products, which help security experts categorize vulnerabilities relevant to their own systems.
A comprehensive overview of vulnerabilities and their impact enables prioritization of defensive measures that need to be taken~\cite{sadlek2022:identification}.
To protect organizations from harmful attacks and assess their security posture, security analysts routinely monitor the NVD for vulnerabilities that are added or updated. 

While vulnerability databases like NVD provide many technical details, 
they do not provide any information on \emph{how} these vulnerabilities may be exploited in real-world attacks. 
The \attack{} knowledge base\footnote{~\url{https://attack.mitre.org}} aims to address this gap by documenting the tactics, techniques, and procedures (TTPs) of adversaries that have been observed in reported attacks~\cite{strom2020:mitre}. 
Understanding how an attacker operates helps formulate a strategy for addressing existing vulnerabilities.
Unfortunately, the NVD does not currently label vulnerabilities with their corresponding TTPs, 
which means that there is no direct connection between the NVD and the \attack{} knowledge base.
As a consequence, assessing the risks associated with a given vulnerability and identifying the most relevant TTPs remains a manual, expertise-driven task -- making the process both time-consuming and prone to various errors and oversights. 

To mitigate this challenge, \citeauthor{mitre2025:mapping} created the CVE Mapping Methodology (\methodology{})~\cite{mitre2025:mapping} which guides the mapping of CVEs to their relevant \attack{} techniques or tactics.
However, while the \methodology{} is inherently valuable in assisting security analysts with manual mapping, the sheer volume of vulnerabilities published each year calls for the development of a more automated solution to keep CVEs aligned with their corresponding \attack{} tactics and techniques.

We present a novel approach that combines the \methodology{} with Large Language Models (LLMs) to automatically map CVEs to \attack{} techniques. 
LLMs have demonstrated strong in-context learning capabilities, enabling them to perform complex tasks with only a few task-specific training examples provided as part of their prompt~\cite{dong2024:survey}. 
This makes in-context learning particularly well-suited for cybersecurity applications, where labeled data are scarce and costly to produce.

\head{Contributions} 
The main contribution of this paper is \approach{} (``Technique Ranking via In-context learning And Guided Extraction''), 
a two-pronged approach that automatically maps CVEs to their corresponding \attack{} techniques. 
We first prompt an LLM using guidance from the \methodology{} to generate a list of relevant \attack{} techniques. 
Next, we instruct the LLM to use in-context learning on existing CVEs and their mapping to \attack{} techniques to generate a second list of relevant techniques. 
Finally, we combine the information from both lists into an overall prediction. 
This hybrid approach strategically combines rule-based mapping from the \methodology{} with data-driven in-context learning based on already labeled CVEs.
We empirically evaluate \approach{} using a MITRE dataset that maps existing CVEs to corresponding \attack{} techniques~\cite{mitre2025:known:zenodo}, and we compare our approach to the current state-of-the-art.
Moreover, we assess performance across two LLMs, the proprietary GPT-4o-mini and the open-weight Llama3.3-70B, and conduct an ablation study to analyze the contributions of various parts of the prompt used for in-context learning.
Our evaluation shows that \approach{} outperforms the state-of-the-art, and the proprietary model GPT-4o-mini outperforms the open-weight model Llama3.3-70B on this task. 
We conclude that LLMs can be used to automatically predict the impact of
cybersecurity vulnerabilities and \approach{} makes the process of mapping
CVEs to \attack{} more efficient. 
We include the various prompts in the Appendix and provide a replication package~\cite{host2025:replication}.

\section{Background \& Related Work}\label{sec:related_work}

\subsection{CVE Mapping Methodology (CMM)}\label{sec:related_work:cve_mappings}\label{sec:cmm}

\begin{figure}[t]
    \centering
    \includegraphics[width=.9\columnwidth]{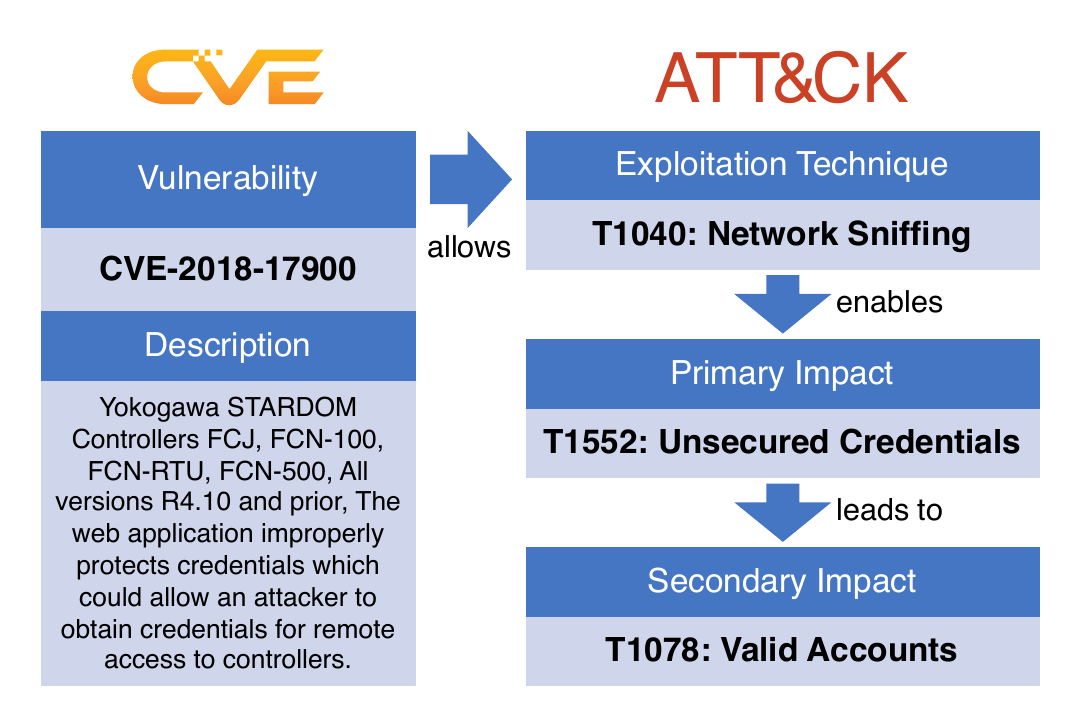}
    \caption{A vulnerability is labeled in three attack steps with associated attack techniques (based on an example in the mapping methodology \cite{mitre2025:mapping}). }
    \label{fig:mapping_methodology}
    \vspace*{-.5ex}
\end{figure}

\noindent
The \methodology{} defines an approach to map CVEs to \attack{} techniques~\cite{mitre2025:mapping}.
These mappings help determine the impact of a vulnerability and how it can be exploited in the real world.
The \methodology{} distinguishes attack techniques related to three attack phases:
First, a vulnerability enables the attacker to use a specific \emph{exploitation technique}.
Second, the exploit grants the attacker certain capabilities, known as the \emph{primary impact}. 
Third, the attacker can gain further capabilities, leading to the \emph{secondary impact}.
These phases, illustrated in \cref{fig:mapping_methodology}, are adapted from the 
\emph{intrusion kill chain} introduced by \citeboth{hutchins2011:intelligencedriven}.
In each phase, the defender can detect and prevent the attack using various strategies.

A vulnerability can be mapped to multiple \attack{} techniques, and a mapping is a one-to-one relationship between a vulnerability and an attack technique, labeled with one of the three possible mapping types (exploitation technique, primary or secondary impact).
The \methodology{} provides five methods to analyze a CVE.
A single method will not necessarily provide all relevant attack techniques across the three mapping types, 
but the methods can be combined to obtain more complete mappings.
The five methods are as follows: 
\begin{enumerate}
\item \label{method:vultype} \textbf{Vulnerability type}: 
    This method supports mapping a CVE based on its vulnerability type. 
    It defines a table with relevant attack techniques (or instructions on how to find them) for 27 of the most common vulnerability types.     
    These vulnerability types are based on CWEs; however, some of them group multiple CWEs to keep the number of alternatives manageable. 
    An example is \textsl{Cleartext Transmission of Sensitive Information} which is mapped to \textsl{T1040: Network Sniffing} as exploitation technique, \textsl{T1552: Unsecured Credentials} as primary impact, and \textsl{T1078: Valid Accounts} as secondary impact. 

\item \label{method:func} \textbf{Functionality}:
    This method defines a table that classifies a CVE by the functionality a successful exploit provides. 
    For example, the ability to \textsl{Delete Files}, which maps to the primary impact \textsl{T1485: Data Destruction}, and secondary impact \textsl{T1499: Endpoint Denial of Service}.

\item \label{method:exploit} \textbf{Exploitation Technique}:
    This method asks the question: ``what steps are needed to exploit this CVE?'', and defines a set of possible answers that are linked to relevant attack techniques.
    For example, if the user needs to visit a malicious website, 
    the exploitation technique can be mapped to \textsl{T1189: Drive-by Compromise}.
    
\item \label{method:affected} \textbf{Affected Object}:
    This method defines a table of objects that may be affected by an exploit, 
    and provides mappings to associated attack techniques. 
    Objects include hardware, firmware, operating systems, applications, services, etc. 
    For example, \textsl{Network-based Application} is mapped to exploitation techniques \textsl{T1040 Network Sniffing}, and \textsl{T1059: Command and Scripting Interpreter}.
    
\item \label{method:tactic} \textbf{Tactic}:
    This method defines a table that associates \attack{} tactics with generic exploitation techniques. 
    Using this method is only recommended when other methods do not provide a relevant answer.
    For example, the tactic \textsl{Lateral Movement} is associated with the technique \textsl{T1210 Exploitation of Remote Services}.    
\end{enumerate}

\subsection{Mining \itattack{} Techniques From Linked Data}\label{sec:linked_data}

\noindent
Cybersecurity data sources, such as CVE, CWE, CAPEC and \attack{}, capture related and at times overlapping concepts. 
These data can be \emph{mined} to uncover mappings from CVEs to attack techniques. 
\citeboth{hemberg2021:linking} investigated this route by creating BRON, a graph database that combines these data sources and unifies the relations between them. 
The graph can be queried, and edges can be traversed as bi-directional links to identify patterns, discover new connections, and gain deeper insights into threats and vulnerabilities.
The authors find that the links are often sparse; for example, not every CVE or CWE is linked to an associated attack technique.
In a follow-up paper~\cite{hemberg2024:enhancements}, the authors extended BRON with additional sources and propose an approach to predict missing links.
However, the focus was on predicting links to neighboring nodes (e.g CWE to CAPEC), and they do not predict multi-step links, such as from CVE to \attack{} techniques.

A clear benefit of establishing such linked data is the reduced cost of labeling.
However, in addition to the sparsity challenge, linking through CWEs -- which are more generic than CVEs -- may lead to a loss of context, for example, regarding attack steps indicated by the CVE description. 
As a consequence, the techniques mined through such links will cover much more than what would be mapped through the \methodology{}.
This aspect is investigated by \citeboth{simonetto:comprehensive}, who unify data sources similar to BRON, but enforce stricter conditions on the CWEs, showing that relations between multiple CWEs may lead to an exploding attack surface.
In one example, they find that BRON maps to 84 different techniques, whereas their more strict approach maps to only two. They also opt for uni-directional linking, which they find has considerable advantages on analysis speed.

\subsection{Automating the Mapping from CVE to \itattack{}}\label{sec:related_work:mapping_cve_attack}

\noindent
\head{Data-driven methods}  
\attack{} distinguishes various levels, and the literature includes approaches for automating the mapping from CVE to \attack{} \emph{tactics}~\cite{ampel2021:linking, branescu2024:automated}, \attack{} \emph{techniques} \cite{grigorescu2022:cve2attck,abdeen2023:smet}, and both \cite{zhang2024:vttllm}. 
Tactics are described at a higher level than techniques, 
for example, \textsl{Initial Access} describes the adversary \emph{goal} of getting into a network.
\emph{How} the adversary enters the network is then described through attack techniques associated with the tactic, for example \textsl{T1566: Phishing}.
Our review will focus on automated mapping from CVEs to \emph{techniques}, 
which is the goal of \approach{}.

\noindent
\citeboth{grigorescu2022:cve2attck} compare multi-labeled supervised approaches with and without data augmentation.
They include classical machine learning (Support Vector Machines and Naive Bayes Classifiers), 
deep learning models trained from scratch, 
and fine-tuning of pre-trained models based on variants of BERT architectures. 
The classical models are used to establish a baseline and help interpret relevant features. 
Not surprisingly, these models do not perform at the same level as the deep learning models, 
particularly the pre-trained BERT-based architectures. 
The authors perform data augmentation using the TextAttack\footnote{~\url{https://textattack.readthedocs.io/en/master}} framework to generate adversarial examples by transforming the input text.
They find that data augmentation boosts the performance of the deep learning models, but not the classical models.
Their two best-performing approaches relied on SciBERT and SecBERT to encode the representations of the CVE text descriptions. %

\citeboth{abdeen2023:smet} propose SMET, which uses Semantic Role Labeling (SRL) and sentence embedding to map CVEs to techniques.
SMET does not need CVEs labeled with attack techniques and considers all 185 techniques that were in \attack{} at the time of the study.
It consists of three steps: 
(1) attack vector extraction through Semantic Role Labeling (SRL), 
(2) attack vector representation using \attack{} BERT, and 
(3) attack technique prediction using Logistic Regression trained on the embeddings produced by \attack{} BERT. 

The attack vector extraction of SMET identifies subject-verb-object structures from the input text using an SRL model from AllenNLP.
SMET is trained on attack vectors extracted from technique descriptions and procedure examples from \attack{}.
An attack vector is extracted if either the words ``adversary,'' ``vulnerability,'' or ``user'' are among the subject words, or the verb is either ``allow,'' ``lead,'' or ``result.'' 
The authors do not discuss why these words were selected, or to what extent they cover the attack vectors found in CVEs.

The authors develop their own \attack{} BERT model which is fine-tuned using the SIAMESE network to optimize the similarity between two attack vectors.
In this context, two attack vectors are considered similar if they either share the same objective or help to achieve the same technique.
The authors find that SMET using their own \attack{} BERT outperforms instances of SMET that use SecBERT or SBERT.

\citeboth{aghaei2023:cvedriven} also apply SRL to this task. However, in contrast to SMET, 
which exclusively uses the attack vectors extracted from CVE descriptions, 
they also consider the surrounding context (the full CVE description) as input to the classification model.
Moreover, they incorporate \emph{attack functionality prediction} as an intermediate step towards predicting the attack techniques. 
This step partially draws on the \methodology{}, as predicted functionalities can be mapped to ATT\&CK techniques using the methodology’s predefined associations (method \Cref{method:func}). 
However, functionality mapping represents only one of the five methods outlined in the \methodology{}, and it does not provide comprehensive coverage of all techniques. 
In contrast, our proposed \approach{} approach leverages the full range of methods from the \methodology, along with labeled examples. 
Moreover, it is the only automated approach that explicitly models the three mapping types from the \methodology{} (exploitation techniques, primary and secondary impacts), providing additional context to each mapping, and offering more complete insight on how a vulnerability is exploited.

\citeboth{adam2022:attack} propose an approach that first maps a CVE to a CWE, and then uses both the \methodology{} and linked data (\Cref{sec:linked_data}) to map from CWE to attack techniques.
Their approach, which is based on \citeboth{aghaei2020:threatzoom}, performs feature extraction on CVE descriptions experimenting with both word embeddings and TF-IDF.
Using the extracted features, they train a set of neural networks to predict CWEs from CVEs.
In addition to approach described above, the authors experiment with associating CVEs to techniques directly using a similarity search approach with Doc2Vec~\cite{le2014:distributed}, however this alternative approach performs significantly worse.

\head{LLMs}
As part of evaluating SMET, \citeboth{abdeen2023:smet} study whether ``one of the most popular chatbots'' can map CVEs to techniques.
They provide this chatbot (left unnamed in the paper) with a CVE description and ask to map it to \attack{} techniques.
They also apply the chatbot as the extraction component in SMET.
In both cases, SMET consistently outperforms the LLM.
The authors experimented with three different prompts, all of them relatively short, 
limited to the CVE description, and two sentences for instructions.

\citeboth{liu2023:not} investigate the use of LLMs for two tasks: (1) mapping from CVE to CWE, and (2) mapping from CVE to \attack{}. 
The authors experiment with \emph{weak} prompts, which are straightforward questions related to the task, and \emph{strong} prompts, which use a chain-of-thought process where they first ask five options for possible answers, and then ask for the final answer based on these. 
The authors consider the performance of the LLM in all experiments to be unsatisfactory, which they attribute to  poor quality of the data, as well as the LLM lacking a conceptual understanding of \attack{}. 

\citeboth{zhang2024:vttllm} experiment with different fine-tuning approaches to make LLMs more domain-specific. Their approach, VTT-LLM, shows a clear improvement compared to \citeboth{liu2023:not}.
The first stage of their two-stage approach uses data from BRON to prompt a generative LLM aiming to capture the links from CVE via CWE and CAPEC to \attack{} techniques and tactics and turn a CVE description into a technical attack description. 
Their best approach uses a template chain, explicitly capturing these links by asking the LLM to ``think step by step'' over each data point.
In the second stage, they use embeddings to associatively match the generated attack description with technique descriptions derived from the \attack{} knowledge base.

Finally, \citeboth{rafiey2024:mapping} explore an idea related to the present paper, describing how they prompt ChatGPT based on the \methodology{} and few-shot examples. 
However, their contributions are difficult to evaluate, as their prompts only define overall mapping goals as the context and include a CVE description in the user prompt, omitting details for the five methods of the \methodology{}, as well as the few-shot examples of labeled CVEs.
The authors mention consolidating the primary and secondary impacts due to the difficulty of establishing the relationship and ordering between them.

\head{Comparative Analysis and Reproduction Studies}
\citeauthor{jaouhari2024:improving} re-implement and compare five related studies~\cite{jaouhari2024:improving}.
In addition, the authors add hyperparameter tuning and data augmentation to each approach. 
They provide an overview of the differences in results but do not compare in more detail.
As such, two challenges remain:
First, the evaluations often employ different metrics that cannot be directly compared.
Second, there is significant variation in the number of attack techniques covered by the different datasets, 
varying from 17 to 52 in the considered literature.
Training and testing the approaches from each study on standard datasets using the same metrics would be interesting for a more detailed comparison.
Finally, although the authors report strong results for their data augmentation, 
closer analysis of their procedure shows that training samples can leak into their test set
because the data are augmented \emph{before} splitting out the test set.

\section{The Hybrid \textsc{Triage} Approach}\label{sec:approach}

\begin{figure}[t]
    \vspace*{.5mm} %
    \centering
    \includegraphics[width=.9\columnwidth]{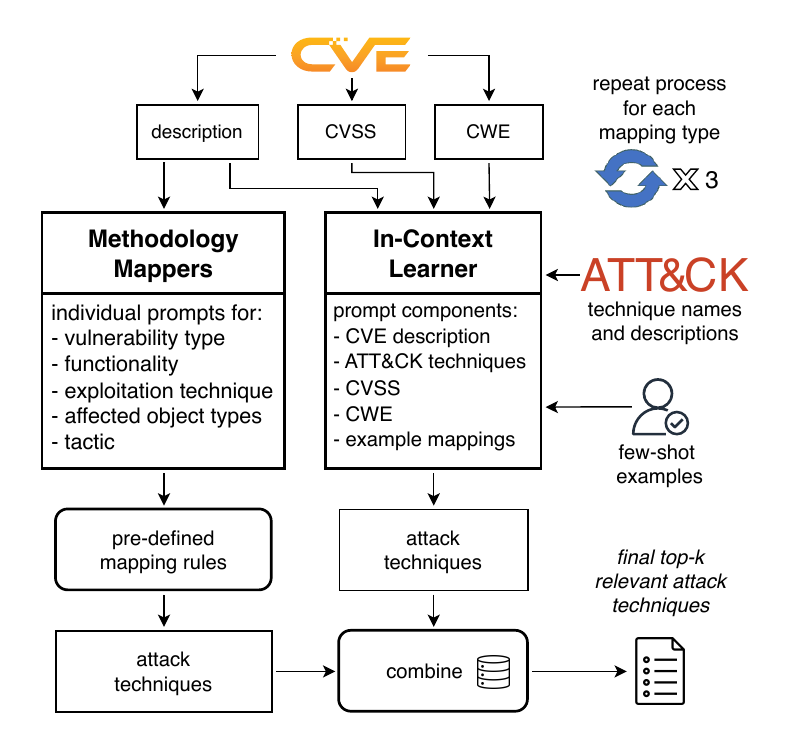}
    \caption{Overview of the \approach{} approach. The figure illustrates how a ranked list of attack techniques is predicted from a given vulnerability.}
    \label{fig:pipeline}
    \vspace*{-.5ex}
\end{figure}

\noindent
Our goal is to take a CVE as input and automatically map it to relevant attack techniques defined in \attack{}.
We propose a hybrid approach with multiple separate LLM invocations to capture the various aspects of each CVE. 
Finally, we combine the output of individual runs to obtain a list of technique predictions per mapping type. 
\approach{} is illustrated in \Cref{fig:pipeline} and consists of two main components, 
the \textsl{Methodology Mappers}, which implement rule-based mapping and prompting following the \methodology{}, 
and the \textsl{In-Context Learner}, which implements many-shot in-context learning based on already labeled CVEs.

\subsection{Methodology Mappers}

\noindent
As input, the model takes a CVE with its corresponding description.
Using a template-based approach, we create prompts for each of the five methods of the \methodology{},
and run these separately through the LLM. 
The prompt templates used for each method are detailed in Appendices \ref{sec:appendix:templates:vulnerability_type} to \ref{sec:appendix:templates:tactic}.

\head{Vulnerability type}
Vulnerability types are closely related to CWEs.
Thus, we can match vulnerability types with corresponding CWE types, 
and use the CWE description as a description of the vulnerability type. 
Based on this list of descriptions, we ask the LLM to identify which vulnerability type matches the CVE description the best.
Although occasionally multiple CWEs can be present in a CVE record, 
the common practice is to map a CVE to a single CWE.
For this reason, we instruct the LLM to provide only one vulnerability type. 
Predicted vulnerability types are then mapped to attack techniques using the \methodology{} table for this method (\cref{method:vultype}).

\head{Functionality}
Our prompt first explains what a functionality is, and lists possible functionality types in the \methodology{}.
Descriptions of functionality types are not included in the \methodology{}, but we reuse the descriptions provided by \citeboth{aghaei2023:cvedriven}.
We allow the LLM to predict multiple functionalities, which are then mapped to attack techniques using the \methodology{} table for this method (\cref{method:func}).

\head{Exploitation Technique}
The \methodology{} defines eight top-level questions for this method, which in two cases are
followed by a multiple-choice sub-question. 
First, we provide the LLM with the top-level questions as ``Yes/No'' questions. 
When relevant, we provide it with the sub-questions and answers to choose from. 
The model can answer ``Yes'' to multiple top-level questions. 
Aggregated answers are mapped to attack techniques using the \methodology{} rules for this method (\cref{method:exploit}).

\head{Affected Object}
The prompt includes a description of affected objects along with possible categories. 
Descriptions of individual object types are not provided in the \methodology{}, 
so we manually created a list of descriptions for the various object types in this method.
We allow the model to predict multiple affected objects which are then mapped to attack techniques using the \methodology{} table for this method (\cref{method:affected}). 

\head{Tactic}
The prompt contains a general definition of a tactic according to the \attack{} knowledge base, and provides brief descriptions of each possible type. 
We allow the model to predict multiple tactics which are mapped to attack techniques following the \methodology{} table for this method (\cref{method:tactic}).

\subsection{In-Context Learner}
\noindent
The \textsl{In-Context Learner} creates attack mappings based on relevant features and in-context examples without any direct reference to the \methodology. 
The prompt was developed iteratively, adding relevant features and refining these based on the response.
The prompt template is detailed in \Cref{sec:appendix:templates:in-context}.
It begins with categorizing an attack in three steps: Exploitation technique, Primary Impact and Secondary Impact. 
Next, we provide all enterprise attack techniques as classes to choose from. 
For each technique, we include the attack names and the full descriptions from \attack{}.
Then, for each target CVE, we include CVSS scores and CWE types in addition to the CVE description.
Next, we include in-context examples from our training set.    %
Each in-context example includes CVE ID, CVE description, attack technique ID, attack name, and mapping type.
The last part of the prompt summarizes the task and the output format, 
in which the LLM is instructed to predict a ranked list of the ten most relevant attack techniques for a given mapping type.

\subsection{Combining Results Into Final Predictions}{\label{sec:approach:final_predictions}}

\noindent
To obtain the final attack technique predictions we combine the results from the \textsl{Methodology Mappers} and the \textsl{In-Context Learner}.
First, from the \textsl{Methodology Mappers}, the intermediate predictions of each method are aggregated in a list.
Next, for each CVE and mapping type, we combine the list of top ten predictions from the \textsl{In-Context Learner} with the aggregated list of \textsl{Methodology Mappers} predictions.

The \textsl{In-Context Learner} is a purely data-driven approach, while the \textsl{Methodology Mappers} mimic a rule-based expert system. 
By combining these two approaches, we aim to get the \emph{best of both worlds} in our final predictions.

\subsection{Research Questions}

\begin{enumerate}[label=\textbf{RQ\arabic*:}, leftmargin=*]
\item To what extent can systematically applying the \methodology{} automate the mapping from CVE to \attack{}?  
\item Is there a performance difference between prompting based on the \methodology{} vs.\ using in-context learning with ground truth labels based on the methodology?
\item How can we combine the two approaches from RQ2, and to what extent does this lead to performance gains?

\item What is the practical relevance of \approach{} in terms of efficacy and quality of the resulting predictions?

\end{enumerate}

\section{Experimental Design}\label{sec:experimental_design}

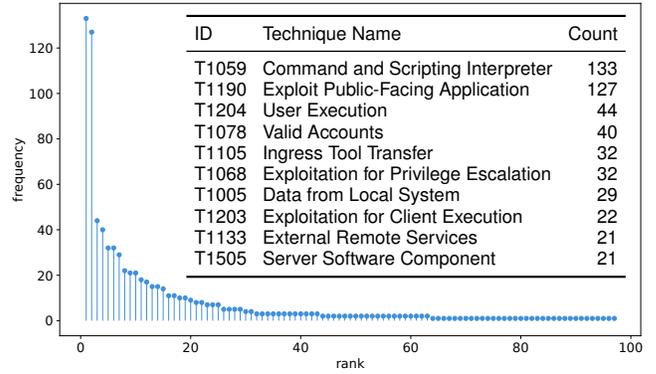
\begin{figure}[t]
    \centering
    \includegraphics[width=\columnwidth]{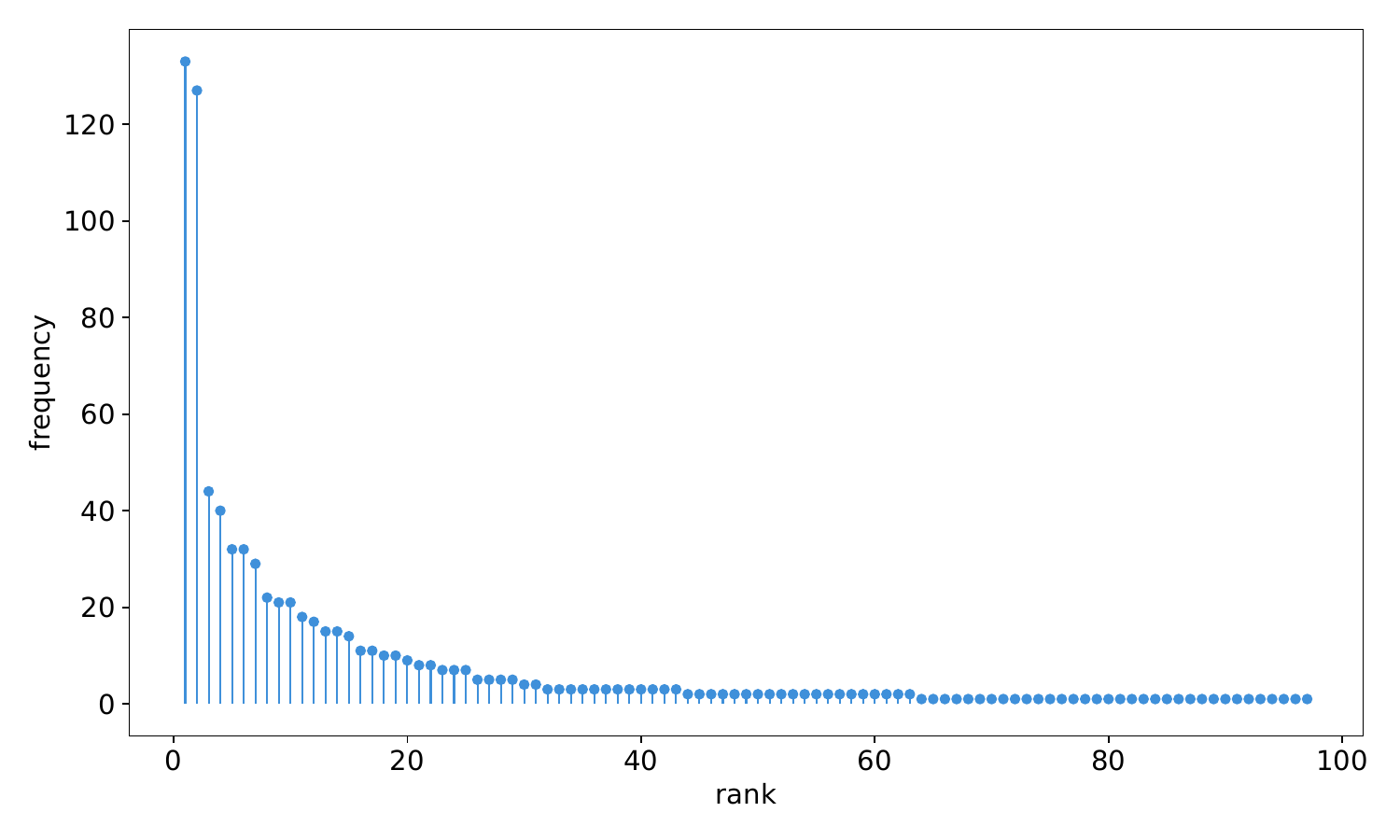}
    \begin{picture}(0,0) %
        \put(-52,152){%
            \raisebox{-\height}{\scriptsize\begin{filecontents*}{tables/data/top_techniques.csv}
attack_id,count,attack_name
T1059,133,Command and Scripting Interpreter
T1190,127,Exploit Public-Facing Application
T1204,44,User Execution
T1078,40,Valid Accounts
T1105,32,Ingress Tool Transfer
T1068,32,Exploitation for Privilege Escalation
T1005,29,Data from Local System
T1203,22,Exploitation for Client Execution
T1133,21,External Remote Services
T1505,21,Server Software Component
T1496,18,Resource Hijacking
T1189,17,Drive-by Compromise
T1486,15,Data Encrypted for Impact
T1574,15,Hijack Execution Flow
T1003,14,OS Credential Dumping
T1136,11,Create Account
T1071,11,Application Layer Protocol
T1566,10,Phishing
T1087,10,Account Discovery
T1202,9,Indirect Command Execution
T1499,8,Endpoint Denial of Service
T1070,8,Indicator Removal
T1498,7,Network Denial of Service
T1082,7,System Information Discovery
T1041,7,Exfiltration Over C2 Channel
T1046,5,Network Service Discovery
T1048,5,Exfiltration Over Alternative Protocol
T1027,5,Obfuscated Files or Information
T1608,5,Stage Capabilities
T1090,4,Proxy
T1552,4,Unsecured Credentials
T1053,3,Scheduled Task/Job
T1036,3,Masquerading
T1210,3,Exploitation of Remote Services
T1021,3,Remote Services
T1221,3,Template Injection
T1485,3,Data Destruction
T1556,3,Modify Authentication Process
T1548,3,Abuse Elevation Control Mechanism
T1573,3,Encrypted Channel
T1557,3,Adversary-in-the-Middle
T1083,3,File and Directory Discovery
T1555,3,Credentials from Password Stores
T1560,2,Archive Collected Data
T1562,2,Impair Defenses
T1482,2,Domain Trust Discovery
T1218,2,System Binary Proxy Execution
T1565,2,Data Manipulation
T1497,2,Virtualization/Sandbox Evasion
T1622,2,Debugger Evasion
T1584,2,Compromise Infrastructure
T1195,2,Supply Chain Compromise
T1112,2,Modify Registry
T1055,2,Process Injection
T1056,2,Input Capture
T1185,2,Browser Session Hijacking
T1140,2,Deobfuscate/Decode Files or Information
T1047,2,Windows Management Instrumentation
T1098,2,Account Manipulation
T1114,2,Email Collection
T1213,2,Data from Information Repositories
T1110,2,Brute Force
T1106,2,Native API
T1588,1,Obtain Capabilities
T1553,1,Subvert Trust Controls
T1567,1,Exfiltration Over Web Service
T1571,1,Non-Standard Port
T1601,1,Modify System Image
T1592,1,Gather Victim Host Information
T1598,1,Phishing for Information
T1569,1,System Services
T1550,1,Use Alternate Authentication Material
T1211,1,Exploitation for Defense Evasion
T1547,1,Boot or Logon Autostart Execution
T1543,1,Create or Modify System Process
T1007,1,System Service Discovery
T1016,1,System Network Configuration Discovery
T1018,1,Remote System Discovery
T1033,1,System Owner/User Discovery
T1037,1,Boot or Logon Initialization Scripts
T1040,1,Network Sniffing
T1049,1,System Network Connections Discovery
T1069,1,Permission Groups Discovery
T1119,1,Automated Collection
T1134,1,Access Token Manipulation
T1212,1,Exploitation for Credential Access
T1217,1,Browser Information Discovery
T1219,1,Remote Access Software
T1222,1,File and Directory Permissions Modification
T1484,1,Domain Policy Modification
T1489,1,Service Stop
T1490,1,Inhibit System Recovery
T1491,1,Defacement
T1530,1,Data from Cloud Storage
T1531,1,Account Access Removal
T1542,1,Pre-OS Boot
T1653,1,Power Settings
\end{filecontents*}
\pgfplotstabletypeset[
                columns={attack_id,attack_name,count},
                font=\sffamily\scriptsize,
                type right aligned=count,
                skip rows between index={10}{99999},
                every head row/.style={
                    output empty row, %
                    before row={
                        \toprule
                        ID  & Technique Name & Count  \\
                    },
                    after row={\midrule},
                },
                string replace*={baz}{qux},
            ]{tables/data/top_techniques.csv}}%
        }    
    \end{picture}
    \vspace*{-5ex} %
    \caption{%
        Rank-frequency curve of \attack{} techniques in the dataset combined with a table showing the top-10 techniques.
    }
  \label{fig:labeldata:full:counts}
  \vspace*{-2ex}
\end{figure}

\subsection{Dataset}\label{sec:experimental_design:datasets}

\noindent
This work builds on the KEV (Known Exploited Vulnerabilities) dataset provided by \citeboth{mitre2025:known:zenodo}.
The dataset contains 296 CVEs labeled with \attack{} techniques, comprising 806 mappings. 
The distribution of individual techniques over these mappings is visualized in \cref{fig:labeldata:full:counts}.
As can be observed, the distribution of mapped techniques is highly imbalanced.
Moreover, exploration of the dataset revealed that a significant number of mappings refer to techniques not mentioned in the \methodology{}, and several techniques included in the \methodology{} do not occur in the dataset.\footnote{~A total of 
14 of 306 exploitation technique mappings,
34 of 256 primary impact mappings, and
58 of 244 secondary impact mappings in the dataset refer to techniques that are \emph{not} in the \methodology.
Conversely, from all techniques \emph{in} the \methodology{}, the dataset uses 11 of 15 exploitation techniques, 16 of 24 primary impacts, and 14 of 18 secondary impacts.
\label{fn:dataset}}
Our interpretation is that the annotator must have relied on external knowledge not included in the \methodology{}.
The methodology does not prohibit this; however, our \textsl{Methodology Mapper} follows the five \methodology{} methods and will predict corresponding techniques.
On the other hand, the \textsl{In-Context Learner} should be able to handle such cases, so we keep these mappings in the dataset.

\Cref{tab:mapping_types_per_CVE} shows how frequent each mapping type is across the CVEs in the dataset.
The table highlights that exploitation technique and primary impact most commonly have one technique while secondary impacts mostly have none which occurs when a mapping type is missing for a CVE.

From the dataset, we extract CVE IDs, attack IDs, attack names, and mapping types.
Moreover, we enrich the data by adding CVE descriptions, CVSS scores, CWE name, and CWE description to each CVE.
Similarly, for each attack technique, we add \attack{} descriptions. 
The 806 mappings in the data correspond to 296 CVEs. 
We split the labeled data in train and test splits with an 80/20 split over the CVEs.

\begin{table}[t]
\centering
\changetabcolsep{4pt}
\caption{Frequency of mapping types per CVE}
\vspace*{-.5ex}
\label{tab:mapping_types_per_CVE}
\pgfplotstableread[col sep=comma,trim cells]{tables/data/mapping_types_per_cve.csv}\data

\pgfplotstabletranspose[
    colnames from=techniques-per-cve,
    columns={techniques-per-cve,exploitation-occurrences,primary-occurrences,secondary-occurrences},
]{\transposed}{\data}

\pgfplotstabletypeset[
    type right aligned/.list={Zero, One, Two, Three, More than three},
    every head row/.style={
        output empty row, %
        before row={ %
            \toprule
             mapping type & none & one & two & three & $>$ three \\
        },
        after row={\midrule},
    },
    string replace*={exploitation-occurrences}{exploitation},
    string replace*={primary-occurrences}{primary},
    string replace*={secondary-occurrences}{secondary},
]{\transposed}
 \vspace*{-.5ex}
\end{table}

\subsection{Evaluation Metrics}\label{sec:experimental_design:evaluation_metrics}

\noindent
In addition to evaluating our final predictions, we also evaluate each method (prompts) of the \textsl{Methodology Mappers} along with various configurations of the \textsl{In-Context Learner}. 
Here we describe the metrics used in these evaluations. 

\head{Unranked Classification}
The \emph{Methodology Mappers} generate \emph{unranked} predictions of how a CVE can be classified into techniques. These are multi-label predictions that cover the three mapping types.
We also consider a variation on our \textsl{In-Context Learner} that generates a similar unranked multi-label classification.
To score these results, we use the standard evaluation metrics for \emph{unranked sets}: precision (P), recall (R), and their harmonic mean F1~\cite[Ch.\ 8.3]{manning2008:introduction}.

We adapt the metrics to multi-label classification by micro-averaging, which places equal weight on individual instances across all classes.
We focus on micro-averaging since our data is imbalanced, and we want to avoid that rare classes have a disproportionate influence on the overall metric value.
A challenge with using micro-averaging on the highly imbalanced data with many classes is that a model that always predicts the majority class can achieve good performance.  
To mitigate this, we also analyze performance over classwise predictions, which are computed by framing the multi-label task as multiple binary classification problems, treating each target technique as positive and all others as negative.

\head{Ranked Results}
Both the \textsl{In-Context Learner} and the final predictions produce \emph{ranked} lists of techniques for each of the three mapping types.
To score these results, we use evaluation metrics that consider not only whether a technique is correct but also its position in the list~\cite[Ch. 8.4]{manning2008:introduction}.

Assume that, for one of the three mapping types, a given CVE $i$ is mapped to a (possibly empty) list $T^*$ of associated techniques in the ground truth (the labeled dataset). 
Let us also assume the model outputs an ordered list $\hat{T}$ of techniques predicted to be associated with the CVE for that mapping type. 
We can then specify a cut-off threshold $k$ and define $\text{Precision}@k$, abbreviated to P@$k$, as the number of correctly predicted techniques in the top-$k$ divided by $k$:
\begin{equation}\label{eq:eval:rank:pre}
\text{P@}k = \frac{|\{t_i \in \hat{T}_{1:k} \mid t_i \in T^*\}|}{k}
\end{equation}
where $\hat{T}_{1:k}$ are the top-$k$ techniques from the ones predicted by the model. Similarly, $\text{Recall@}k$ or R@$k$ is defined as:

\begin{equation}\label{eq:eval:rank:rec}
\text{R@}k = \frac{|\{t_i \in \hat{T}_{1:k} \mid t_i \in T^*\}|}{|T^*|}
\end{equation}

\noindent
For $\text{Average Precision}$, or $\text{AP}$, we compute $\text{Precision@}i$ for all possible values of $i$:
\begin{equation}\label{eq:eval:rank:ap}
\text{AP} = \frac{1}{n} \sum_{i=1}^{o} \text{P@}i \cdot \mathbb{1}(\hat{t}_i \in T^*)
\end{equation}
where $n = | T^* |$, $o = | \hat{T} |$ and $\mathbb{1}$ denotes the indicator function. 

Finally, we compute the Mean Average Precision (or $\text{MAP}$) as the mean of all individual $\text{AP}$ scores across all labeled examples in the ground truth dataset, denoted as Q:
\begin{equation}\label{eq:eval:rank:map}
\text{MAP} = \frac{1}{Q}\sum_{q=1}^Q \text{AP}_q
\end{equation}
The $\text{MAP}$ score measures both whether or not a prediction is correct but in addition also considers quality of the ranking.

\subsection{Technical Details}

\noindent
We describe here some implementation choices that were not discussed when describing the overall approach in \Cref{sec:approach}.

\head{Lifting Sub-Techniques}
In the \textsl{In-Context Learner}, we lift all sub-techniques to their top-level parent technique to reduce the number of classes that the model must handle and shorten the list of examples in the prompt.

\head{Combining Lists of Ranked Predictions}
We use the following strategy for aggregating predictions from the \textsl{Methodology Mappers} and \textsl{In-Context Learner}:
(1) we want to avoid overwriting high-ranked predictions from the \textsl{In-Context Learner} with lower-ranked predictions from the \textsl{Methodology Mappers};
(2) we want to keep ``None'' as prediction from the \textsl{In-Context Learner} when it appears.
Thus, we merge \textsl{Methodology Mapper}-generated predictions into \textsl{In-Context Learner} ones in descending order, starting from the last position.
We overwrite any lower-ranked \textsl{In-Context Learner} prediction with novel predictions from the \textsl{Methodology Mappers}, except when there is a ``None'' value at the corresponding position. 
This way, we get the best from the \textsl{In-Context Learner}, augmented by novel additions from the \textsl{Methodology Mappers}.

\subsection{Ablation Study}

\noindent 
To investigate how different components of the \textsl{In-Context-Learner} prompt influence the performance,
we run a set of ablations over 100 randomly selected CVEs, and focus on the following four features:
(1) \emph{Attack descriptions} provide extra context to the options the model should select from;
(2) \emph{CVSS metrics} include information on how the vulnerability may be exploited and what aspects of the system are affected;
(3) \emph{CWE types} are closely related to vulnerability types and may indicate attack techniques, cf. the \methodology{} (\cref{method:vultype});
(4) \emph{In-context examples} enable in-context learning from existing mappings.
However, they are generally expensive to create and increase the context window. 

\subsection{Enabling a Comparison between \itapproach{} and SMET}\label{sec:experimental_design:compare_smet}

\noindent
We chose to compare our results to SMET~\cite{abdeen2023:smet}, which also produces a ranked list of techniques.
This requires a few additional steps.
First, we need to collect SMET results for the same dataset as we use.
To collect the results, we can reuse the models previously trained by SMET's authors, as these depend on \attack{}, not on the evaluation dataset.

The next step comes from the fact that SMET does not distinguish between the three mapping types from the \methodology{}.
To compare \approach{}, which does distinguish these mapping types, we perform post-processing of our results to merge the three mapping types into a single category. 
Since these are ranked lists, the merging is done as follows:
The top-ranked technique is ``popped'' from the list of exploitation techniques and added to the results; 
then we do the same for the top of the primary impact list, and then the top of the secondary impact list. 
We repeat this process until all the lists are empty. 

\section{Results and Discussion}\label{sec:results_and_discussion}

\begin{table}[t]
\centering
\caption{Unranked classification with GPT-4o-mini.
The upper part shows individual Methodology Mappers (MM), the middle part shows their union, and the lower part shows the In-Context Learner. Scores are micro-averaged.}
\label{tab:exact_matching:method}
\pgfplotstableread[col sep=comma,trim cells]{tables/data/exact_matching_methodology.csv}\data

\pgfplotstabletypeset[
    columns/mapping-method/.style={string type},
    row predicate/.code={%
        \pgfplotstablegetelem{\pgfplotstablerow}{mapping-method}\of{\data}%
    	\edef\cell{\pgfplotsretval}%
    	\def\matchA{two-step-approach}%
    	\def\matchB{methodology-combine}%
    	\ifx\cell\matchA \pgfplotstableuserowfalse%
            \else \ifx\cell\matchB \pgfplotstableuserowfalse%
            \fi\fi%
    },
    columns={mapping-type,mapping-method,train F1 score,train precision,train recall,test F1 score,test precision,test recall},
    suppress duplicates=mapping-type,
    replace mapping types=mapping-type,
    replace mapping methods=mapping-method,
    type siunitx/.list={train F1 score,train precision,train recall,test F1 score,test precision,test recall},
    every head row/.style={%
        output empty row, %
        before row={%
            \toprule%
            \headingcell{mapping\\type}  & \headingcell{mapping\\method} & \multicolumn{3}{c}{train} & \multicolumn{3}{c}{test} \\%
            \cmidrule(lr){3-5} \cmidrule(lr){6-8}%
            &    &  {F1} & {P} & {R}         &  {F1} & {P} & {R} \\%
        },
        after row={\midrule},
    },
    every last row/.style={after row=\midrule},
    string replace*={baz}{qux},
]{\data}\newline%

\pgfplotstabletypeset[%
    columns/mapping-method/.style={string type},
    row predicate/.code={%
        \pgfplotstablegetelem{\pgfplotstablerow}{mapping-method}\of{\data}%
    	\edef\cell{\pgfplotsretval}%
    	\def\match{methodology-combine}%
    	\ifx\cell\match \else \pgfplotstableuserowfalse\fi%
    },
    columns={mapping-type,mapping-method,train F1 score,train precision,train recall,test F1 score,test precision,test recall},
    suppress duplicates=mapping-type,
    replace mapping types=mapping-type,
    replace mapping methods=mapping-method,
    type siunitx/.list={train F1 score,train precision,train recall,test F1 score,test precision,test recall},
    every head row/.style={%
        before row={},
        output empty row, %
    },
    every last row/.style={after row=\midrule},
    string replace*={baz}{qux},
]{\data}\newline%

\pgfplotstabletypeset[%
    columns/mapping-method/.style={string type},
    row predicate/.code={%
        \pgfplotstablegetelem{\pgfplotstablerow}{mapping-method}\of{\data}%
    	\edef\cell{\pgfplotsretval}%
    	\def\match{two-step-approach}%
    	\ifx\cell\match \else \pgfplotstableuserowfalse\fi%
    },
    columns={mapping-type,mapping-method,train F1 score,train precision,train recall,test F1 score,test precision,test recall},
    suppress duplicates=mapping-type,
    replace mapping types=mapping-type,
    replace mapping methods=mapping-method,
    type siunitx/.list={train F1 score,train precision,train recall,test F1 score,test precision,test recall},
    every head row/.style={%
        before row={},
        output empty row, %
    },
    string replace*={baz}{qux},
]{\data}%
 \vspace*{-.5ex}
\end{table}

\subsection{Assessing the Individual Components of \itapproach{}}\label{sec:results:unranked}

\noindent
We first explore how well individual mapping methods compare, considering all \textsl{Methodology Mappers}, and the \textsl{In-Context Learner} (ignoring the ranking for now).
The upper part of \Cref{tab:exact_matching:method} shows the results for individual \textsl{Methodology Mappers} (observe that not all are relevant for each mapping type); 
the middle part shows them combined, taking the union of all answers per mapping type; 
and the lower part shows the \textsl{In-Context Learner}.
Focusing on the F1 scores, we see that the \textsl{In-Context Learner} consistently outperforms the other methods. 
The \textsl{vulnerability type} method is far better at predicting primary impact than exploitation technique or secondary impact.
As expected, the \textsl{exploitation technique} method is the best among \emph{Methodology Mappers} for predicting exploitation techniques.
The \textsl{functionality} method performs poorly for both primary and secondary impacts on both data splits.
(and does not predict exploitation techniques cf. the \methodology{}).
The combination of all \textsl{Methodology Mappers} performs well in recall, but poorly in precision. 
This is in part due to the noise introduced by the poorly performing methods (precision increases when they are removed). 
An underlying cause here is that a considerable number of techniques mentioned in the \methodology{} are not used in the dataset, so every prediction of these creates a false positive (see also \cref{sec:experimental_design:datasets}).

\subsection{Comparing Two Different LLMs for Powering \itapproach{}}

\noindent
Moving towards our hybrid approach, we change from unranked to ranked predictions.
As a baseline, we will first assess the performance of the \textsl{In-Context Learner} by itself.
Next, we will assess the performance of our hybrid combination of the \textsl{Methodology Mappers} with the \textsl{In-Context Learner}.
Considering the poor performance of various \textsl{Methodology Mappers} in the previous section, only the following are included in our hybrid:
(1) vulnerability type,
(2) exploitation technique,
(3) affected object.
We will also use this opportunity to compare the performance of two LLMs, the proprietary \textsl{GPT-4o-mini} and the open-weight \textsl{Llama3.3-70B}.

The upper part of \Cref{tab:ranking_approach:models} shows the performance of both LLMs using the \textsl{In-Context Learner} alone. 
Considering the MAP scores, \textsl{GPT-4o-mini} consistently outperforms \textsl{Llama3.3-70B} across all mapping types and data splits. 
Looking at recall, \textsl{GPT-4o-mini} always performs better than \textsl{Llama3.3-70B}, except Recall@5 for exploitation technique and Recall@10 for primary impact, where they score equally.

The lower part of \Cref{tab:ranking_approach:models} shows the performance of \approach{}.
Taking into account the MAP scores, \textsl{GPT-4o-mini} again outperforms \textsl{Llama3.3-70B} in all mapping types and data splits. 
We see that the performance related to $\text{Recall@10}$ is almost the same between the two, indicating that the hybrid approach evens out the initial differences related to exploitation technique.
This suggests that \textsl{Methodology Mappers} extract relevant exploitation techniques that are not identified by the \textsl{In-Context Learner}.
For primary or secondary impact, we do not see a similar effect.
Moreover, for secondary impact, we see a large difference when comparing $\text{Recall@10}$ with $\text{Recall@5}$ in \cref{tab:ranking_approach:models}, which means that many of the extracted techniques are not among the top 5 ranked predictions. 
This is also reflected in the low MAP scores for secondary impact, indicating that the model is poor at ranking the correct techniques.
We also note that $\text{Recall@10}$ increases significantly for secondary impact when comparing the training and test sets for both models. 
This could be related to the fact that the test set has far fewer unique attack techniques (18) compared to the training set (71), when considering the secondary impacts.
Rare techniques are typically more difficult to predict, which we will examine in more detail in the next section.
At a high level, the results from \Cref{tab:ranking_approach:models} indicate that exploitation techniques are the easiest types to predict for the model, followed by primary impacts and finally secondary impacts.

\begin{table}[t]
\centering
\changetabcolsep{1.4pt}
\caption{{GPT-4o-mini} vs.\ {Llama3.3-70B} on the In-Context Learner (ICL, upper part)
and the complete \approach{} approach (lower part).}
\label{tab:ranking_approach:models}

\pgfplotstableread[col sep=comma,trim cells]{tables/data/ranking_approach_models.csv}\data

\begin{adjustbox}{width=\columnwidth}%
\pgfplotstabletypeset[%
    columns/model/.style={string type},
    row predicate/.code={%
        \pgfplotstablegetelem{\pgfplotstablerow}{model}\of{\data}%
    	\edef\cell{\pgfplotsretval}%
    	\def\matchA{gpt-4o-mini-combined}%
    	\def\matchB{llama3.3-70b-combined}%
    	\ifx\cell\matchA \pgfplotstableuserowfalse%
            \else \ifx\cell\matchB \pgfplotstableuserowfalse%
            \fi\fi%
    },
    suppress duplicates=mapping-type,
    columns={mapping-type,model,train MAP,train Recall-at-ten,train Recall-at-five,test MAP,test Recall-at-ten,test Recall-at-five},
    type siunitx/.list={train MAP,train Recall-at-ten,train Recall-at-five,test MAP,test Recall-at-ten,test Recall-at-five},
    replace mapping types=mapping-type,
    replace models=model,
    every head row/.style={%
        output empty row, %
        before row={%
            \toprule%
            \headingcell{mapping\\type} & \headingcell{mapping\\method} & \multicolumn{3}{c}{train} & \multicolumn{3}{c}{test} \\%
            \cmidrule(lr){3-5} \cmidrule(lr){6-8}%
                & \phantom{hybrid (GPT-4o-mini)} & MAP & {R@10} & {R@5} & MAP & {R@10} & {R@5} \\%
        },
        after row={\midrule},
    },
    every last row/.style={after row=\midrule},
    string replace*={baz}{qux},
]{\data}
\end{adjustbox}

\begin{adjustbox}{width=\columnwidth}%
\pgfplotstabletypeset[%
    columns/model/.style={string type},
    row predicate/.code={%
        \pgfplotstablegetelem{\pgfplotstablerow}{model}\of{\data}%
    	\edef\cell{\pgfplotsretval}%
    	\def\matchA{gpt-4o-mini-combined}%
    	\def\matchB{llama3.3-70b-combined}%
    	\ifx\cell\matchA%
            \else \ifx\cell\matchB%
                  \else \pgfplotstableuserowfalse\fi%
            \fi%
    },
    suppress duplicates=mapping-type,
    columns={mapping-type,model,train MAP,train Recall-at-ten,train Recall-at-five,test MAP,test Recall-at-ten,test Recall-at-five},
    type siunitx/.list={train MAP,train Recall-at-ten,train Recall-at-five,test MAP,test Recall-at-ten,test Recall-at-five},
    replace mapping types=mapping-type,
    replace models=model,
    every head row/.style={%
        output empty row, %
        before row={%
            \phantom{exploitation} & \phantom{hybrid (GPT-4o-mini)} & \phantom{MAP} & \phantom{{R@10}} & \phantom{{R@5}} & \phantom{{MAP}} & \phantom{{R@10}} & \phantom{{R@5}} \\[-2.2ex]%
        },
        after row={},
    },
    string replace*={baz}{qux},
]{\data}%
\end{adjustbox}

 \vspace*{-.5ex}
\end{table}

\subsection{Ablation Study and Analysis of Classwise Predictions}\label{sec:results_and_discussion:ablation_and_classwise}

\noindent
In addition to \cref{sec:results:unranked}, which assesses the contributions of \approach{} components, we conduct an ablation study on the \textsl{In-Context Learner}, gradually removing features from the prompt.
As shown in \cref{tab:ranking_approach:ablations}, the CVSS and CWE features have a limited impact on the various mapping types.
For primary impact, including CVSS even reduces recall.
Omitting attack descriptions (\textsl{A}) decreases all three metrics for primary and secondary impact.
For exploitation technique, omitting attack descriptions increases the MAP by one percentage point, while $\text{Recall@5}$ and $\text{Recall@10}$ decrease.
A clear pattern is that exploitation techniques mapping benefits from more in-context examples.
Reducing from 235 to 30 examples reduces performance by 7 to 10 percentage points across the metrics. 
An even larger performance drop is seen when removing all examples.
The secondary impact is less influenced by removing features, which is likely due to the generally lower performance for this mapping type. 

\begin{table}[t]
\centering
\changetabcolsep{4pt}
\caption{{In-Context Learner} running ablations with 100 CVEs\\(A in features are the attack technique descriptions;\\\#demo are the number of demonstrations included).}
\label{tab:ranking_approach:ablations}
\begin{filecontents*}{tables/data/ranking_approach_ablations.csv}
mapping-type,features,num-demonstrations,MAP,Recall-at-ten,Recall-at-five
exploitation-technique,ACVSSCWE,235,0.55,0.76,0.71
exploitation-technique,ACVSS,235,0.57,0.76,0.71
exploitation-technique,A,235,0.58,0.76,0.74
exploitation-technique,,235,0.59,0.75,0.67
exploitation-technique,,30,0.52,0.65,0.58
exploitation-technique,,0,0.27,0.42,0.34
primary-impact,ACVSSCWE,235,0.42,0.64,0.46
primary-impact,ACVSS,235,0.42,0.63,0.49
primary-impact,A,235,0.42,0.69,0.50
primary-impact,,235,0.42,0.64,0.46
primary-impact,,30,0.43,0.63,0.48
primary-impact,,0,0.19,0.42,0.22
secondary-impact,ACVSSCWE,235,0.13,0.61,0.14
secondary-impact,ACVSS,235,0.11,0.61,0.13
secondary-impact,A,235,0.12,0.59,0.15
secondary-impact,,235,0.11,0.55,0.12
secondary-impact,,30,0.11,0.56,0.09
secondary-impact,,0,0.07,0.48,0.01
\end{filecontents*}
\pgfplotstabletypeset[
    suppress duplicates=mapping-type,
    replace mapping types=mapping-type,
    replace features=features,
    columns={mapping-type,features,num-demonstrations,MAP,Recall-at-ten,Recall-at-five},
    type siunitx/.list={MAP,Recall-at-ten,Recall-at-five},
    /columns/num-demonstrations/.style={column type=r},
    every head row/.style={
        output empty row, %
        before row={
            \toprule
             mapping type & features & {\#demo} & {MAP} &  {R@10} &  {R@5} \\ %
        },
        after row={\midrule},
    },
    columns/num-demonstrations/.style={
        column type=r,
        string type,
    },
    string replace*={foobar}{qux},
]{tables/data/ranking_approach_ablations.csv}

\end{table}

\begin{figure}[t]
    \centering
    \includegraphics[width=\columnwidth,]{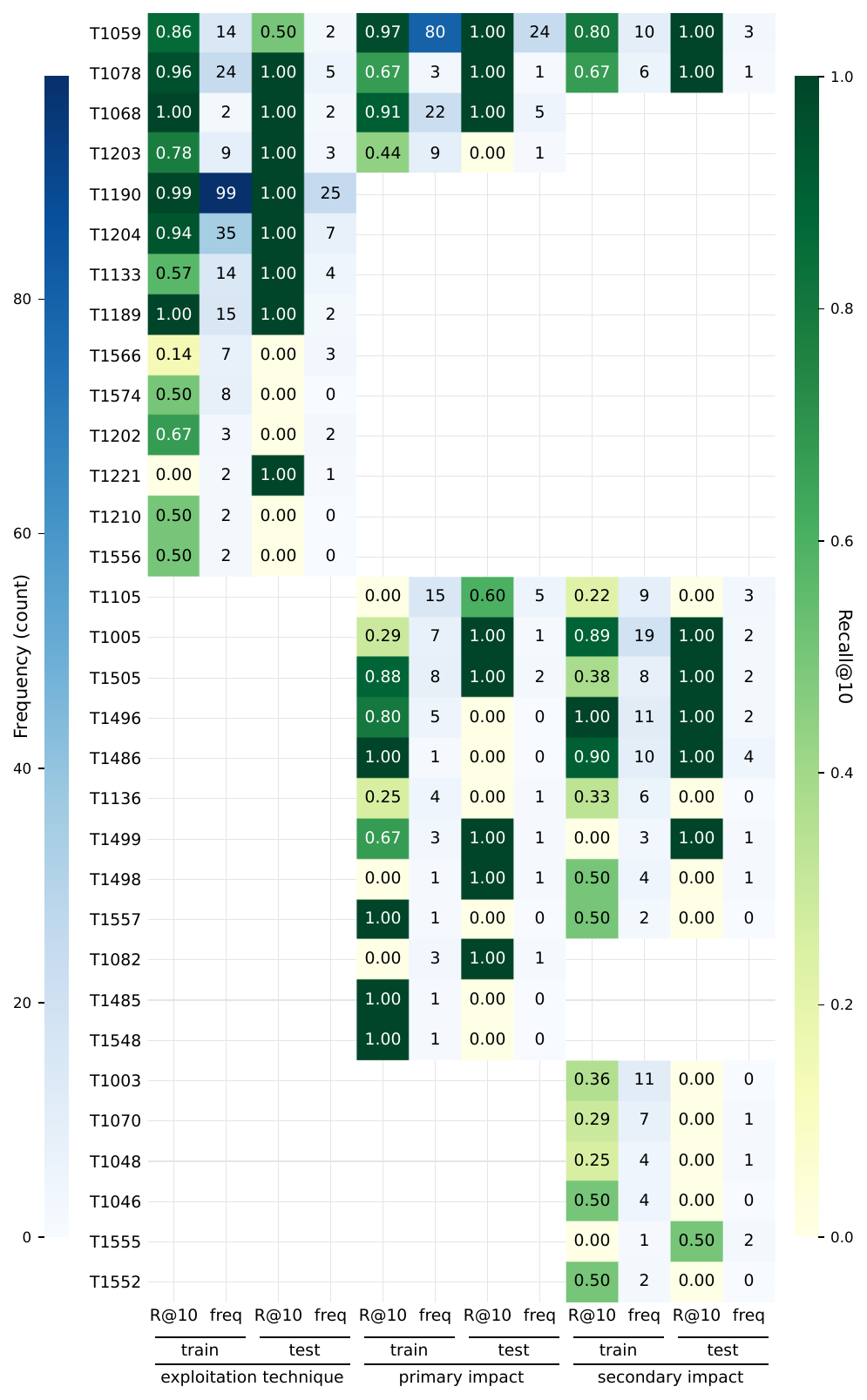}
    \vspace*{-3ex}
    \caption{Classwise performance of \approach{} on each of the three mapping types. The green cells indicate R@10, the blue ones frequency.}
    \label{fig:heatmap}
\end{figure}

The heatmap in \Cref{fig:heatmap} zooms in on the classwise performance for predicting individual \attack{} techniques. 
Note that the figure only includes techniques that have at least one correct prediction among any of the mapping types.
The chart shows large variations in how well individual techniques are predicted across the mapping types and data splits, ranging from perfect scores to no correct predictions. 

\subsection{Performance comparison of SMET and \itapproach{}}

\noindent
When comparing our approach with SMET, we consider CVE descriptions mapped to techniques, without mapping types. 
Moreover, we have seen in our earlier results that secondary impact techniques are more difficult to predict than primary impact and exploitation technique. 
Therefore, we explore two scenarios which, respectively, include and exclude those secondary impacts from the ground truth and results.

In \Cref{tab:compare_smet}, we compare SMET with \approach{} for the two scenarios.
We see that our hybrid approach outperforms SMET across all metrics for both configurations.
We also see in \Cref{tab:compare_smet}, that performance for both approaches improves when secondary impacts are excluded. 

\begin{table}[t]
\centering
\changetabcolsep{4pt}
\caption{Performance comparison of SMET and \approach{}\\ (\approach{}* indicates adapted for uncategorized predictions)}
\label{tab:compare_smet}
\begin{filecontents*}{tables/data/compare_smet.csv}
row,mapping-method,exclude-secondary-impact,train MAP,train Recall-at-ten,train Recall-at-five,test MAP,test Recall-at-ten,test Recall-at-five
0,smet,False,0.08,0.16,0.07,0.20,0.38,0.24
1,ours*,False,0.52,0.62,0.52,0.57,0.70,0.62
2,smet,True,0.20,0.43,0.26,0.21,0.43,0.28
3,ours*,True,0.59,0.77,0.69,0.60,0.72,0.70
\end{filecontents*}
\pgfplotstabletypeset[
    columns={mapping-method,exclude-secondary-impact,train MAP,train Recall-at-ten,train Recall-at-five,test MAP,test Recall-at-ten,test Recall-at-five},
    type siunitx/.list={train MAP,train Recall-at-ten,train Recall-at-five,test MAP,test Recall-at-ten,test Recall-at-five},
    every head row/.style={
        output empty row, %
        before row={
            \toprule
            \headingcell{mapping\\method}  & \headingcell{secondary\\impacts} & \multicolumn{3}{c}{train} & \multicolumn{3}{c}{test} \\ 
            \cmidrule(lr){3-5} \cmidrule(lr){6-8}   
                                           &                                  &  MAP &  {R@10} &  {R@5}   &  MAP &  {R@10} &  {R@5} \\
        },
        after row={\midrule},
    },
    string replace*={False}{included},
    string replace*={True}{excluded},
    string replace*={smet}{SMET},
    string replace*={ours*}{\approach{}*},
]{tables/data/compare_smet.csv}

\end{table}

\subsection{Qualitative Analysis}\label{sec:results_and_discussion:qualitative_analysis}

\noindent
For our qualitative analysis, we randomly selected eight CVEs, 
and found that the predicted mappings were correct for five CVEs, 
while the remaining three had one or more mispredictions.
In addition, we observed that the ground truth for
five CVEs had empty values for the secondary impact, 
and two CVEs had empty values for both the primary and the secondary impact. 
Due to space limitations, we present four of eight analyzed CVEs in this paper, which were selected as follows:
First, we include the three CVEs with mispredictions to investigate how our approach missed the target. 
Analysis of these CVEs is presented in \Cref{cve:CVE-2018-4939}, \ref{cve:CVE-2021-40539}, and \ref{cve:CVE-2022-34713}.
Then, to explore the empty values in the ground truth in more detail, 
we include one CVE with empty values for both primary and secondary impact (\Cref{cve:CVE-2021-21975}).

Each analysis follows the same pattern:
We start by analyzing the differences between the ground truth and the predictions of \approach{}.
Next, we consider to what extent the ground truth labels follow common patterns in the data.
In the end, we manually re-evaluate the CVE description using \methodology{} to better understand potential differences between our predictions and the ground truth. 

Looking at \Cref{cve:CVE-2018-4939,cve:CVE-2021-40539}, we see that the first one has no secondary impact, even when the description directly mentions a vulnerability type with a secondary impact. 
The second one has as much as twelve secondary impacts, but these are hard to trace back to the \methodology{}, and are sparse in the training data.
In \cref{cve:CVE-2022-34713}, we see a case where the ground truth has a primary impact that cannot be derived through the \methodology.
Also noteworthy is \Cref{cve:CVE-2021-21975}, where \approach{} correctly identifies  \emph{plausible} primary and secondary impacts (following \methodology{} rules) that are missing from the ground truth.

\begin{cveanalysis}
  {CVE-2018-4939}
  {Adobe ColdFusion Update 5 and earlier versions, ColdFusion 11 Update 13 and earlier versions have an exploitable Deserialization of Untrusted Data vulnerability. Successful exploitation could lead to arbitrary code execution.}
  {T1203: Exploitation for Client Execution}
  {T1190: Exploit Public-Facing Application, T1133: External Remote Services}
  {}
   The primary impact T1133 is not found by our approach.
   This may be caused by the fact that T1133 most commonly appears as exploitation technique in the ground truth, and this CVE is the only case with T1133 as primary impact.
   Our approach includes T1190 as exploitation technique instead of primary impact. 
   T1190 is most commonly used as exploitation technique in the ground truth,
   but three times it occurs as primary impact. 
   \textsl{Deserialization of Untrusted Data} is mentioned in the CVE description, 
   and is a vulnerability type in the \methodology{} that points to T1059 as primary impact, which is not among the ground truth labels. 
\end{cveanalysis}

\begin{cveanalysis}
  {CVE-2021-40539}
  {Zoho ManageEngine ADSelfService Plus version 6113 and prior is vulnerable to REST API authentication bypass with resultant remote code execution.}
  {T1190: Exploit Public-Facing Application}
  {T1505: Server Software Component}
  { T1573: Encrypted Channel, T1560: Archive Collected Data, T1087: Account Discovery, T1070: Indicator Removal, T1047: Windows Management Instrumentation, T1003: OS Credential Dumping, T1136: Create Account, T1218: System Binary Proxy Execution, T1003: OS Credential Dumping, T1140: Deobfuscate/Decode Files or Information, T1027: Obfuscated Files or Information, T1505: Server Software Component}
Our approach correctly predicts the exploitation technique T1190.
It misses T1505 in the primary impact, but includes it in the secondary impact. 
Interestingly, the ground truth for this CVE includes as many as 12 secondary impact techniques, indicating that there are several different paths open to the attacker.
Among these 12, only T1505 is included in our predictions. 
 Some of the secondary impacts are quite rare in the training data, but all of them occur at least twice, which means there is always at least one in-context example with the same technique available. 
 Earlier, we saw that secondary impact predictions already have the lowest performance; a single or just a few in-context examples may not be enough to predict this mapping type. 
 Finally, when manually mapping the CVE description using \methodology{}, we cannot find answers for the primary and secondary impacts.
\end{cveanalysis}

\begin{cveanalysis}
  {CVE-2022-34713}
  {Microsoft Windows Support Diagnostic Tool (MSDT) Remote Code Execution Vulnerability}
  {T1566: Phishing}
  {T1204: User Execution}
  {T1059: Command and Scripting Interpreter}
Our approach does not predict T1566: Phishing as exploitation technique. 
The first and the second ranked exploitation technique are instead T1204: User Execution
and T1190: Exploit Public-Facing Application. 
 T1566 appears seven times as exploitation technique in the training data so there are at least some in-context examples available. 
Using the \methodology{}, \textsl{Operating System} among \emph{affected object types} with T1574: Hijack Execution Flow seems relevant, the attack technique is also included among the predictions.
Moving on to the primary impact, T1204 is not included among the primary impact predictions where T1059 is ranked at top. 
T1204 is used as primary impact two times meaning there is a single in-context example available. 
Using the \methodology{}, we are not able to map the CVE description to a primary impact.
Regarding the secondary impact, our approach predicts the correct technique T1059 as the fifth ranked technique.
\end{cveanalysis}

\begin{cveanalysis}
{CVE-2021-21975}
{Server Side Request Forgery in vRealize Operations Manager API (CVE-2021-21975) prior to 8.4 may allow a malicious actor with network access to the vRealize Operations Manager API can perform a Server Side Request Forgery attack to steal administrative credentials.}
{T1190: Exploit Public-Facing Application}
{}
{}
Our approach correctly identifies T1190 as an exploitation technique. 
The ground truth has no primary and secondary impacts, which is surprising considering that the term \textsl{Server Side Request Forgery} in the description is a
vulnerability type in the \methodology{} pointing to 
T1133: External Remote Services as exploitation technique, 
T1090: Proxy as primary impact, and both 
T1135: Network Share Discovery and 
T1005: Data from Local System as secondary impacts.
Other potential relevant attack techniques predicted by the In-Context Learner include 
primary impacts 
T1555: Credentials from Password Stores,
T1078: Valid Accounts, and 
T1068: Exploitation for Privilege Escalation, 
and secondary impacts 
T1555, T1078, and T1552: Unsecured Credentials.
\end{cveanalysis}

\subsection{Addressing our Research Questions}

\newcommand{\RQ}[2]{\smallskip\noindent\textbf{RQ#1:} \emph{#2}\newline}

\vspace{-1ex}%
\RQ{1}{To what extent can systematically applying the CMM
automate the mapping from CVE to \itattack{}?}
To address the first research question, we assess the \emph{Methodology Mappers} in our approach, individually prompting each method of the \methodology{}. 
Although there are large variations in performance across mapping types and mapping methods, the results in \Cref{tab:exact_matching:method} suggest that using individual components in \emph{Methodology Mappers} is not sufficient to reach good performance. 
The combination of all methods yields better results, but still falls short in performance.
As mentioned when discussing the dataset (\Cref{sec:experimental_design:datasets}), several mappings are related to techniques not present in the methodology, and this is most prevalent for primary and secondary impacts. 

\RQ{2}{Is there a performance difference between prompting based on the \methodology{} vs.\ using in-context learning with ground truth labels based on the methodology?}
The answer here is a resounding yes. \Cref{tab:exact_matching:method} shows that the \textsl{In-Context Learner} results in higher F1 performance than any of the \textsl{Methodology Mappers} individually or all combined.
Furthermore, the \textsl{In-Context Learner} can provide additional predictions through a ranked approach providing a list of most likely answers.
We consider the \textsl{In-Context Learner} the most impactful component of our approach, given the demonstrated improvement in performance, and the fact that a considerable number of annotations do not strictly follow the \methodology{}.

\RQ{3}{How can we combine the two approaches from RQ2,
and to what extent does this lead to performance gains?}
We experimented with two architectures for our hybrid approach:
(1) The first includes the output from \textsl{Methodology Mappers} in the prompt of the \textsl{In-Context Learner} with in-context examples. 
(2) The second uses the two-pronged approach from \cref{fig:pipeline}, invoking the \textsl{In-Context Learner} for each mapping type.
The first approach ``confused'' the LLM in the sense that the answer of the \emph{Methodology Mappers} took precedence, and no other techniques were added based on the in-context examples.
The second approach shows a performance increase for the exploitation technique but not for the primary or the secondary impact (\Cref{tab:ranking_approach:models}). 

\RQ{4}{What is the practical relevance of \itapproach{} in terms of efficacy and quality of the resulting predictions?}
There are three aspects to answering this question. 
First, we compare our work with SMET, the current state-of-the-art in automated mapping. 
There is a clear improvement in predictive performance for \approach{} compared to SMET (\cref{tab:compare_smet}).
Moreover, our approach provides additional cybersecurity context, by classifying into \methodology{} mapping types.

Next, our qualitative analysis found that \approach{} included the correct mappings for five of eight CVEs.
We observed that even if the CVE directly mentions a \textsl{Vulnerability Type} present in the \methodology{}, the annotator may have selected other attack techniques.
In this case (CVE-2021-21975, \cref{cve:CVE-2021-21975}), \approach{} predicted \emph{plausible} impacts following the \methodology{}, but these were missing from the ground truth. 
Our qualitative analysis supports our quantitative analysis that predicting secondary impacts is a difficult task for the LLMs. The lack of data to learn from is a considerable factor here.

The last aspect to consider is what it costs to use \approach{} to analyze a CVE and predict mappings for all three mapping types, and how long it takes?
The combined number of tokens to map a CVE using \approach{} is in the order of 400k input tokens, divided in three prompts for the \textsl{Methodology Mappers} and three for the \textsl{In-Context Learner}.
Sequentially, the total process takes on average 50 seconds for GPT-4o-mini (OpenAI API) and 2m50s for Llama3.3-70B (Lambda AI API) and produces 2k output tokens.\footnote{~Parallelization speedups depend on user rate limits; we get around 50\%.}
Thus, the overall \approach{} mapping costs per CVE currently correspond to US\$0.07 for GPT-4o-mini and US\$0.05 for Llama3.3-70B.

Comparing again with SMET, run-times or costs are unfortunately not reported by its authors, 
and since we do not train its models, we cannot compute these costs ourselves.
However, when we run inference with SMET on an AMD EPYC 7413 CPU with 2.85 GHz, 
the processing time averages around 0.6 seconds per CVE.
So, purely looking at run times, SMET outperforms \approach{}, although as we have seen, 
it does so at a lower predictive performance.

Based on these findings, we argue that \approach{} has practical relevance, where the \textsl{Methodology Mappers} select specific techniques when there is a clear path in the methodology, and the \textsl{In-Context Learner} provides the flexibility needed to fill in the gaps in the methodology. 
Moreover, using \approach{} to support analysts makes the process of mapping CVEs to \attack{} more efficient, at low cost and runtime overhead.

\subsection{Threats to Validity}

\noindent
\head{Internal validity}
\emph{Data leakage} is a threat and comes in two forms: 
(1) The LLMs may have seen the mapping of the CVEs. 
The dataset's release after the cut-off dates for the LLMs\footnote{~Dataset: 2025-03-12; Cut-off dates are resp. Oct 2023 for GPT-4o-mini, released 2024-07-18, and Dec 2023 for Llama3.3-70B, released 2024-12-06.} mostly mitigates this, but mapped examples might have been online before.
(2) Many CVE descriptions literally include the vulnerability type, e.g., ``SQL Injection'',  which can lead the \textsl{In-Context Learner} to matching examples. We do not consider this a threat in \approach{} since the \textsl{Methodology Mappers} follow \methodology{} rules that do the exact same thing.

\head{External validity} 
One threat is \emph{generalizability beyond our ground truth}.
Performance may differ on non-KEV CVEs, and in other domains.
The lack of more labeled data makes it difficult to mitigate this threat.
There is also the threat of \emph{provider and model dependency.}
We evaluate two LLMs, the proprietary model performed best, but new capable models emerge frequently, reducing this dependency.

\head{Construct validity}
\cref{sec:experimental_design:datasets} discussed the threat of \emph{ground-truth mapping mismatch.}
We mitigate by
(i) using both ranked and unranked metrics, (ii) analyzing coverage per technique group, and (iii) qualitatively inspecting disagreements to separate genuine errors from plausible alternatives.

\head{Conclusion validity} 
\attack{} evolves, and deprecated or merged techniques can yield \emph{ambiguity in labeling}. 
Expert mapping at sub-technique levels is also more subjective. 
To mitigate, we lift sub-techniques to their top-level technique. 

\subsection{Ethical considerations}

\noindent   
We acknowledge that \approach{} has a measurable \emph{dual-use risk}, 
considering that adversaries could use the predictions to prioritize exploitation paths or craft more effective payloads. 
Such dual-use risks are common for security-related tooling, e.g., a malware detector can be misused to assess sufficient obfuscation of malicious code.
In the case of \approach{}, the risk is offset by substantial defensive benefits: 
by predicting the most likely \attack{} techniques for a vulnerability, 
it helps defenders prioritize patches, focus threat-hunting efforts, and map mitigations to realistic attack paths. 
As such, we believe that the benefits of providing defenders with capable tooling far outweigh the risks of misuse by attackers.

\section{Concluding Remarks}

\noindent
This paper presented a variety of ways LLMs can be prompted to map CVEs to \attack{} techniques.  
We conclude that a hybrid combination of prompting with expert rules derived from the \methodology{} and in-context learning yields the best performance, with in-context learning contributing most to the overall results. 
Our proposed approach, \approach{}, can automatically map a vulnerability to a list of potential \attack{} techniques, ranked by their likelihood, and it outperforms the current state-of-the-art in automated, ranked, mapping of CVEs.

Our ablation studies show that the number of in-context examples plays an important role in the performance. 
Moreover, our qualitative analysis reveals that, even when a CVE description explicitly mentions a vulnerability type associated with known attack techniques, the annotators may assign different techniques -- or none at all. 
However, \approach{} was still able to predict \emph{plausible} primary and secondary impacts (cf. the \methodology{}), showing the benefit of the hybrid approach. 

We do see that the approach underperforms in predicting secondary impacts, which we believe is in part due to not having enough labeled training samples 
of this type, but more importantly, because the CVE descriptions lack details that allow reasoning this deep into the \methodology{} intrusion kill chain.
This highlights the complexity of the mapping task and the limitations of relying solely on surface-level cues.

\head{Future Work}
For future work, we see a clear need for further standardization of datasets and annotation guidelines. 
In addition, we are interested in comparing our work to VTT-LLM \cite{zhang2024:vttllm}. 
A direct comparison is not possible, but one way to proceed is to make our approach uncategorized (like in \Cref{sec:experimental_design:compare_smet}) and evaluate it unranked.
In addition, the datasets used by both approaches would need to be aligned.
Finally, we would like to explore the use of reasoning models in \approach{}, examine to what extent LLMs can help analyze threat reports, and investigate using knowledge graphs to structure and integrate existing knowledge into this process.

\section*{Data Availability}
\noindent
To support open science and allow for replication and verification of our work, a replication package with prompts, scripts, and results is made available via Zenodo~\cite{host2025:replication}.

\section*{Acknowledgements}
\noindent
This work is supported by the Research Council of Norway through the secureIT project (\#288787).
The empirical evaluation made use of the Experimental Infrastructure for Exploration of Exascale Computing (eX3), financially supported by the Research Council of Norway under contract \#270053. 

\printbibliography 

@INPROCEEDINGS{sadlek2022:identification,
  AUTHOR = {Sadlek, Lukáš and Čeleda, Pavel and Tovarňák, Daniel},
  BOOKTITLE = {NOMS 2022-2022 IEEE/IFIP Network Operations\ \&\ Management Symp.},
  DATE = {2022-04},
  DOI = {10.1109/NOMS54207.2022.9789803},
  ISSN = {2374-9709},
  LANGID = {english},
  PAGES = {1--6},
  TITLE = {Identification of Attack Paths Using Kill Chain and Attack Graphs},
  URLDATE = {2025-08-07},
}

@REPORT{strom2020:mitre,
  AUTHOR = {Strom, Blake E. and Applebaum, Andy and Miller, Doug P. and Nickels, Kathryn C. and Pennington, Adam G. and Thomas, Cody B.},
  INSTITUTION = {MITRE},
  DATE = {2020},
  LANGID = {english},
  NUMBER = {MP180360R1},
  PAGES = {1--46},
  TITLE = {{{MITRE ATT}}\&{{CK}}: Design and Philosophy},
}

@MISC{mitre2025:mapping,
  AUTHOR = {{MITRE}},
  HOWPUBLISHED = {\textsc{Url:} \url{https://web.archive.org/web/20250804111103/https://center-for-threat-informed-defense.github.io/mappings-explorer/about/methodology/cve-methodology/}},
  TITLE = {{{CVE Mapping Methodology}}},
  URLDATE = {2025-08-04},
}

@INPROCEEDINGS{dong2024:survey,
  AUTHOR = {Dong, Qingxiu and Li, Lei and Dai, Damai and Zheng, Ce and Ma, Jingyuan and Li, Rui and Xia, Heming and Xu, Jingjing and Wu, Zhiyong and Chang, Baobao and Sun, Xu and Li, Lei and Sui, Zhifang},
  LOCATION = {Miami, Florida, USA},
  PUBLISHER = {ACL},
  BOOKTITLE = {Conf.\ Emp.\ Methods in Nat.\ Lang.\ Proc.},
  DATE = {2024-11},
  DOI = {10.18653/v1/2024.emnlp-main.64},
  LANGID = {english},
  PAGES = {1107--1128},
  TITLE = {A Survey on {{In-context}} Learning},
  URLDATE = {2025-07-30},
}

@MISC{mitre2025:known:zenodo,
  AUTHOR = {{MITRE}},
  PUBLISHER = {Zenodo},
  DATE = {2025-08},
  DOI = {10.5281/zenodo.16747173},
  LANGID = {english},
  NOTE = {Version 03.12.2025},
  TITLE = {Known Exploited Vulnerabilities Mapped to {{ATT}}\&{{CK}} Techniques},
  URLDATE = {2025-08-05},
}

@MISC{host2025:replication,
  AUTHOR = {H{ø}st, Anders M{ø}lmen and Lison, Pierre and Moonen, Leon},
  DATE = {2025-10},
  DOI = {10.5281/zenodo.17341503},
  HOWPUBLISHED = {Zenodo},
  TITLE = {Replication {{Package}} for "{{A Systematic Approach}} to {{Predict}} the {{Impact}} of {{Cybersecurity Vulnerabilities Using LLMs}}"},
}

@INPROCEEDINGS{hutchins2011:intelligencedriven,
  AUTHOR = {Hutchins, Eric M and Cloppert, Michael J and Amin, Rohan M},
  BOOKTITLE = {Int'l Conf.\ I-Warfare\ \&\ Security},
  DATE = {2011},
  LANGID = {english},
  TITLE = {Intelligence-{{Driven Computer Network Defense Informed}} by {{Analysis}} of {{Adversary Campaigns}} and {{Intrusion Kill Chains}}},
}

@MISC{hemberg2021:linking,
  AUTHOR = {Hemberg, Erik and Kelly, Jonathan and {Shlapentokh-Rothman}, Michal and Reinstadler, Bryn and Xu, Katherine and Rutar, Nick and O'Reilly, Una-May},
  PUBLISHER = {arXiv},
  DATE = {2021-02},
  EPRINT = {2010.00533},
  EPRINTCLASS = {cs},
  EPRINTTYPE = {arXiv},
  LANGID = {english},
  NUMBER = {arXiv:2010.00533},
  TITLE = {Linking {{Threat Tactics}}, {{Techniques}}, and {{Patterns}} with {{Defensive Weaknesses}}, {{Vulnerabilities}} and {{Affected Platform Configurations}} for {{Cyber Hunting}}},
  URLDATE = {2024-01-26},
}

@ARTICLE{hemberg2024:enhancements,
  AUTHOR = {Hemberg, Erik and Turner, Matthew J. and Rutar, Nick and O'Reilly, Una-May},
  DATE = {2024-03},
  DOI = {10.1145/3615668},
  JOURNALTITLE = {Digital Threats},
  NUMBER = {1},
  PAGES = {8:1--8:33},
  TITLE = {Enhancements to {{Threat}}, {{Vulnerability}}, and {{Mitigation Knowledge}} for {{Cyber Analytics}}, {{Hunting}}, and {{Simulations}}},
  URLDATE = {2024-08-19},
  VOLUME = {5},
}

@INPROCEEDINGS{simonetto:comprehensive,
  AUTHOR = {Simonetto, Stefano and Bosch, Peter},
  BOOKTITLE = {1st Int'l Conf.\ Nat.\ Lang.\ Proc.\ \&\ AI for Cyber Security},
  DATE = {2024-07},
  LANGID = {english},
  PAGES = {32--41},
  TITLE = {Comprehensive Threat Analysis and Systematic Mapping of {{CVEs}} to {{MITRE}} Framework},
}

@INPROCEEDINGS{ampel2021:linking,
  AUTHOR = {{Benjamin Ampel} and {Sagar Samtani} and {Steven Ullman} and {Hsinchun Chen}},
  LOCATION = {Virtual Event Singapore},
  PUBLISHER = {ACM},
  BOOKTITLE = {ACM SIGKDD Conf.\ Knowledge Discovery \& Data Mining},
  DATE = {2021-08},
  DOI = {10.1145/3447548.3469450},
  ISBN = {978-1-4503-8332-5},
  LANGID = {english},
  PAGES = {4153--4154},
  SHORTTITLE = {{{ACM KDD AI4Cyber}}},
  TITLE = {Linking {{Common Vulnerabilities}} and {{Exposures}} to the {{MITRE ATT}}\&{{CK Framework}}: {{A Self-Distillation Approach}}},
  URLDATE = {2024-06-24},
}

@ARTICLE{branescu2024:automated,
  AUTHOR = {Branescu, Ioana and Grigorescu, Octavian and Dascalu, Mihai},
  PUBLISHER = {Multidisciplinary Digital Publishing Institute},
  DATE = {2024-04},
  DOI = {10.3390/info15040214},
  ISSN = {2078-2489},
  JOURNALTITLE = {Information},
  LANGID = {english},
  NUMBER = {4},
  PAGES = {214},
  TITLE = {Automated {{Mapping}} of {{Common Vulnerabilities}} and {{Exposures}} to {{MITRE ATT}}\&{{CK Tactics}}},
  URLDATE = {2024-10-02},
  VOLUME = {15},
}

@ARTICLE{grigorescu2022:cve2attck,
  AUTHOR = {Grigorescu, Octavian and Nica, Andreea and Dascalu, Mihai and Rughinis, Razvan},
  PUBLISHER = {Multidisciplinary Digital Publishing Institute},
  DATE = {2022-09},
  DOI = {10.3390/a15090314},
  ISSN = {1999-4893},
  JOURNALTITLE = {Algorithms},
  LANGID = {english},
  NUMBER = {9},
  PAGES = {314},
  SHORTTITLE = {{{CVE2ATT}}\&{{CK}}},
  TITLE = {{{CVE2ATT}}\&{{CK}}: {{BERT-Based Mapping}} of {{CVEs}} to {{MITRE ATT}}\&{{CK Techniques}}},
  URLDATE = {2023-01-19},
  VOLUME = {15},
}

@INPROCEEDINGS{abdeen2023:smet,
  AUTHOR = {Abdeen, Basel and {Al-Shaer}, Ehab and Singhal, Anoop and Khan, Latifur and Hamlen, Kevin},
  LOCATION = {Cham},
  PUBLISHER = {Springer},
  BOOKTITLE = {Data\ \&\ Applic.\ Security\ \&\ Privacy XXXVII},
  DATE = {2023},
  DOI = {10.1007/978-3-031-37586-6_15},
  ISBN = {978-3-031-37586-6},
  LANGID = {english},
  PAGES = {243--260},
  SERIES = {Lecture {{Notes}} in {{Computer Science}}},
  SHORTTITLE = {{{SMET}}},
  TITLE = {{{SMET}}: {{Semantic Mapping}} of~{{CVE}} to~{{ATT}}\&{{CK}} and~{{Its Application}} to~{{Cybersecurity}}},
}

@ARTICLE{zhang2024:vttllm,
  AUTHOR = {Zhang, Chenhui and Wang, Le and Fan, Dunqiu and Zhu, Junyi and Zhou, Tang and Zeng, Liyi and Li, Zhaohua},
  PUBLISHER = {Multidisciplinary Digital Publishing Institute},
  DATE = {2024-01},
  DOI = {10.3390/math12091286},
  ISSN = {2227-7390},
  JOURNALTITLE = {Mathematics},
  LANGID = {english},
  NUMBER = {9},
  PAGES = {1286},
  SHORTTITLE = {{{VTT-LLM}}},
  TITLE = {{{VTT-LLM}}: {{Advancing Vulnerability-to-Tactic-and-Technique Mapping}} through {{Fine-Tuning}} of {{Large Language Model}}},
  URLDATE = {2024-10-02},
  VOLUME = {12},
}

@MISC{aghaei2023:cvedriven,
  AUTHOR = {Aghaei, Ehsan and {Al-Shaer}, Ehab},
  PUBLISHER = {arXiv},
  DATE = {2023-09},
  EPRINT = {2309.02785},
  EPRINTCLASS = {cs},
  EPRINTTYPE = {arXiv},
  LANGID = {english},
  NUMBER = {arXiv:2309.02785},
  TITLE = {{{CVE-driven Attack Technique Prediction}} with {{Semantic Information Extraction}} and a {{Domain-specific Language Model}}},
  URLDATE = {2024-01-30},
}

@MISC{adam2022:attack,
  AUTHOR = {Adam, Constantin and Bulut, Muhammed Fatih and Sow, Daby and Ocepek, Steven and Bedell, Chris and Ngweta, Lilian},
  PUBLISHER = {arXiv},
  DATE = {2022-06},
  DOI = {10.48550/arXiv.2206.11171},
  EPRINT = {2206.11171},
  EPRINTCLASS = {cs},
  EPRINTTYPE = {arXiv},
  NUMBER = {arXiv:2206.11171},
  TITLE = {Attack {{Techniques}} and {{Threat Identification}} for {{Vulnerabilities}}},
  URLDATE = {2023-11-15},
}

@INPROCEEDINGS{aghaei2020:threatzoom,
  AUTHOR = {Aghaei, Ehsan and Shadid, Waseem and {Al-Shaer}, Ehab},
  EDITOR = {Park, Noseong and Sun, Kun and Foresti, Sara and Butler, Kevin and Saxena, Nitesh},
  LOCATION = {Cham},
  PUBLISHER = {Springer International Publishing},
  BOOKTITLE = {Security\ \&\ Privacy in Communication Networks},
  DATE = {2020},
  DOI = {10.1007/978-3-030-63086-7_2},
  ISBN = {978-3-030-63086-7},
  LANGID = {english},
  PAGES = {23--41},
  SERIES = {Lecture {{Notes}} of the {{Institute}} for {{Computer Sciences}}, {{Social Informatics}} and {{Telecommunications Engineering}}},
  SHORTTITLE = {{{ThreatZoom}}},
  TITLE = {{{ThreatZoom}}: {{Hierarchical Neural Network}} for {{CVEs}} to {{CWEs Classification}}},
}

@MISC{le2014:distributed,
  AUTHOR = {Le, Quoc V. and Mikolov, Tomas},
  PUBLISHER = {arXiv},
  DATE = {2014-05},
  DOI = {10.48550/arXiv.1405.4053},
  EPRINT = {1405.4053},
  EPRINTCLASS = {cs},
  EPRINTTYPE = {arXiv},
  NUMBER = {arXiv:1405.4053},
  TITLE = {Distributed {{Representations}} of {{Sentences}} and {{Documents}}},
  URLDATE = {2025-08-07},
}

@INPROCEEDINGS{liu2023:not,
  AUTHOR = {Liu, Xin and Tan, Yuan and Xiao, Zhenghang and Zhuge, Jianwei and Zhou, Rui},
  EDITOR = {Rogers, Anna and {Boyd-Graber}, Jordan and Okazaki, Naoaki},
  LOCATION = {Toronto, Canada},
  PUBLISHER = {ACL},
  BOOKTITLE = {Findings of the Association for Computational Linguistics: ACL 2023},
  DATE = {2023-07},
  DOI = {10.18653/v1/2023.findings-acl.229},
  PAGES = {3724--3731},
  SHORTTITLE = {Not {{The End}} of {{Story}}},
  TITLE = {Not {{The End}} of {{Story}}: {{An Evaluation}} of {{ChatGPT-Driven Vulnerability Description Mappings}}},
  URLDATE = {2024-08-06},
}

@MISC{rafiey2024:mapping,
  AUTHOR = {Rafiey, Pasha and Namadchian, Amin},
  PUBLISHER = {ResearchSquare},
  DATE = {2024-05},
  DOI = {10.21203/rs.3.rs-4341401/v2},
  EPRINTTYPE = {ResearchSquare},
  LANGID = {english},
  NOTE = {ResearchSquare preprint},
  TITLE = {Mapping {{Vulnerability Description}} to {{MITRE ATT}}\&{{CK Framework}} by {{LLM}}},
  URLDATE = {2024-06-21},
}

@INPROCEEDINGS{jaouhari2024:improving,
  AUTHOR = {Jaouhari, Saad EL and Tamani, Nouredine and Jacob, Rohan ISAAC},
  BOOKTITLE = {IEEE Annual Comp., Softw.,\ \&\ Applic.\ Conf.},
  DATE = {2024-07},
  DOI = {10.1109/COMPSAC61105.2024.00392},
  ISSN = {2836-3795},
  PAGES = {2442--2447},
  TITLE = {Improving {{ML-Based Solutions}} for {{Linking}} of {{CVE}} to {{MITRE ATT}}\&{{CK Techniques}}},
  URLDATE = {2025-01-02},
}

@BOOK{manning2008:introduction,
  AUTHOR = {Manning, Christopher D. and Raghavan, Prabhakar and Schütze, Hinrich},
  LOCATION = {Cambridge},
  PUBLISHER = {Cambridge University Press},
  DATE = {2008},
  ISBN = {978-0-521-86571-5},
  LANGID = {english},
  NOTE = {Available online at \url{https://www-nlp.stanford.edu/IR-book/}},
  TITLE = {Introduction to Information Retrieval},
}

\def\appendixname{Appendix - Prompt Templates}
\appendix \footnotesize
\crefalias{section}{appendix}
\crefalias{subsection}{appendix}

\subsection{Vulnerability Type}\label{sec:appendix:templates:vulnerability_type}

\begin{lstlisting}[style=prompt]
The following vulnerability types (keys) with 
corresponding descriptions (values) exist: 
{  'Deserialization of Untrusted Data': <CWE-DESCRIPTION>,
   <...>
}
Given a new CVE with description:
<CVE-DESCRIPTION>
Which vulnerability type does this CVE map to? Provide
only the vulnerability type. If no vulnerability type 
applies, answer with 'N/A'
\end{lstlisting}

\subsection{Functionality}\label{sec:appendix:templates:functionality}

\noindent
This prompt is repeated for each of the 14 functionalities:
\begin{lstlisting}[style=prompt]
Determine whether or not an attacker gains access to the 
following functionality: 
<FUNCTIONALITY-DESCRIPTION>
by exploiting this vulnerability:
<CVE-DESCRIPTION-a>
Examples of other CVEs that give an attacker access to
this functionality include:
<CVE-DESCRIPTION-i, ..., CVE-DESCRIPTION-k>
Examples of other CVEs that do not give an attacker
access to this functionality include:
<CVE-DESCRIPTION-m, ..., DESCRIPTION-n>
Answer with 'YES' or 'NO'.
\end{lstlisting}

\subsection{Exploitation Technique}

\noindent
The top-level questions prompt is always used completely:
\begin{lstlisting}[style=prompt]
Did the user execute a malicious file?
...
Did the attacker 'sniff' unencrypted network traffic?
\end{lstlisting}

\noindent
The follow-up questions depend on whether certain top-level questions had a positive answer (cf. \methodology{}~\cite{mitre2025:mapping}): 
\begin{lstlisting}[style=prompt]
Where did this file come from?
Answer with one of these alternatives: 
[ "A malicious link", "An email", "A third-party service", 
  "Removable media", "Other" ]
\end{lstlisting}

\subsection{Affected Objects}

\begin{lstlisting}[style=prompt]
An exploitation technique is the method (technique) used 
to exploit the vulnerability. The types of objects that 
are affected by an exploitation technique include 
software, hardware, firmware, product, application, or 
code. Your goal is to predict the affected object relevant
to the exploitation of the vulnerability. There should 
only be one affected object. In the remainder, you will 
first receive the vulnerability to label, then an 
overview of affected objects to use as labels.
The vulnerability to be labeled has the following 
description: 
<CVE-DESCRIPTION>
The JSON string below describes the affected objects to 
choose from:
[ {
      "affected_object": "Internet-facing Host/System",
      "description": "An Internet-facing host is any 
          system that provides a doorway from the open 
          Internet into your infrastructure.",
      "examples": ["webserver", "website", "database", 
          "service"],
      "exploitation_technique": ["T1190", "T1211"]
  }
  ...
]
Based on this information, what is the affected object? 
Select one of the following:
   [ "Internet-facing Host/System", ..., 
     "External Remote Service" ]
\end{lstlisting}

\subsection{Tactic}\label{sec:appendix:templates:tactic}

\begin{lstlisting}[style=prompt]
A tactic is the adversary's tactical goal: the reason for
performing an action. For example, an adversary may want
to achieve credential access. The task is to map from CVE 
to tactic. 
Please map a CVE with the following description: 
<CVE-DESCRIPTION>
to a tactic among the following tactics with corresponding
descriptions: 
<TACTIC-NAMES: TACTIC-DESCRIPTIONS>
\end{lstlisting}

\subsection{In-Context Learner}\label{sec:appendix:templates:in-context}

\noindent
The context variable includes:
\begin{lstlisting}[style=prompt]
You are a cybersecurity analyst specialized in applying 
MITRE's ATT&CK Framework to label vulnerability 
descriptions.
\end{lstlisting}

\bigskip
(... continued on next page to avoid a page break in the prompt ...)
\newpage

\noindent
This prompt is repeated for each of the three mapping types:
\begin{lstlisting}[style=prompt]
An attack consists of three steps which corresponds to the
following mapping types: 
- Exploitation Technique - the method (technique) used to
  exploit the vulnerability
- Primary Impact - the initial benefit (impact) gained
  through exploitation of the vulnerability
- Secondary Impact - what the adversary can do by gaining
  the benefit of the primary impact
Given a vulnerability, your task is to determine the
relevant attack techniques of type exploitation technique.
You should output the top 10 most relevant attack
techniques in descending order. An empty label indicating
no relevant attack technique can be included among the 
top 10 where appropriate. In the remainder, you will first
receive the vulnerability to label, then an overview of
attack techniques to use as labels. Finally, you receive
examples of correctly labeled vulnerabilities.
The vulnerability to be labeled has the following
description:
<CVE-DESCRIPTION>
The data in CSV format below describes the attack
techniques to choose from:
----
attack_id,attack_name
T1047,Windows Management Instrumentation
<...>
---- 
The following CWE applies to this vulnerability:
<CWE-ID>: <CWE-DESCRIPTION>
The CVE has the following CVSS features:
----
- cvssV2.accessVector: <CVSS-ACCESS-VECTOR>
  <...>
----
The following is a JSON with examples. Each entry
corresponds to a vulnerability which has a description
and associated attack techniques by mapping type: 
----
{
   <CVE-ID>: {
      "description": <CVE_DESCRIPTION>,
      "attack_techniques":{
         "exploitation_technique": <TECHNIQUE_DETAILS>,
         "primary_impact": <TECHNIQUE_DETAILS>,
         "secondary_impact": <TECHNIQUE_DETAILS>
      }  
   }  
}
<...>
----
Your task is to determine the relevant attack techniques
of type <MAPPING-TYPE> for the given vulnerability. 
Provide a ranked list of either ten technique IDs or nine 
technique IDs and one 'None' value indicating the empty
label. The 'None' value should be ranked similarly as 
the other techniques when applicable. 
Here is an example of predicted output: ['T1068', None, 
'T1168', 'T1290', 'T1078', 'T1180', 'T1010', 'T1435', 
'T1320', 'T1100'].
The output should have the same format used in the 
example, and an extra explanation should not be included.
\end{lstlisting}

 \end{document}